%% file: ricci-2d-cdt-final.tex
\documentclass[12pt]{article}

\input{incs.tex}
\usepackage[export]{adjustbox}
\usepackage{hyperref}
\usepackage[title]{appendix}

\input{defs.tex}
\graphicspath{{figures/}}

\newcommand{\dbar}{\bar{d}}

\newcommand{\delmax}{\delta_\textrm{max}}

\begin{document}


\begin{center}
${}$\\
\vspace{100pt}
{ \Large \bf Quantum Flatness \\
\vspace{15pt}in Two-Dimensional CDT Quantum Gravity
}

\vspace{46pt}

{\sl J. Brunekreef}
and {\sl R. Loll}

\vspace{24pt}
{\footnotesize

Institute for Mathematics, Astrophysics and Particle Physics, Radboud University \\ 
Heyendaalseweg 135, 6525 AJ Nijmegen, The Netherlands.\\ 
\vspace{12pt}
{email: j.brunekreef@science.ru.nl, r.loll@science.ru.nl}\\
}
\vspace{24pt}

\end{center}

\vspace{0.8cm}

\begin{center}
{\bf Abstract}
\end{center}

\noindent Flatness -- the absence of spacetime curvature -- is a well-understood property of macroscopic, classical spacetimes in general relativity. The same
cannot be said about the concepts of curvature and flatness in nonperturbative quantum gravity, where the microscopic structure
of spacetime is not describable in terms of small fluctuations around a fixed background geometry. An interesting case are two-dimensional
models of quantum gravity, which lack a classical limit and therefore are maximally ``quantum". We investigate the recently 
introduced quantum Ricci curvature in CDT quantum gravity on a two-dimensional torus, whose quantum geometry could be expected to behave like a flat space on
suitably coarse-grained scales. On the basis of Monte Carlo simulations we have performed, 
with system sizes of up to 600.000 building blocks, this does not seem to be the case. 
Instead, we find a scale-independent ``quantum flatness", without an obvious classical analogue. As part of our study, we develop a criterion that allows 
us to distinguish between local and global, topological properties of the toroidal quantum system.

\vspace{12pt}
\noindent


\newpage

\section{Introduction}
In classical general relativity, spacetime is characterised in terms of its curvature properties. The complete local information is
contained in the Riemann curvature tensor, whose fourth-rank nature reflects the complexity of curvature. The challenge of {\it quantum} gravity
is to describe the structure and dynamics of spacetime in a quantum realm. 
Until recently, it was unclear whether a meaningful notion of curvature can be defined in quantum gravity beyond perturbation theory,
in the sense of a finite, renormalized quantum curvature operator that remains well-defined in a (near-)Planckian regime. 
This has changed with the advent of the quantum Ricci curvature, introduced in \cite{qrc1,qrc2}. It was subsequently implemented 
 in full, four-dimensional quantum gravity, formulated in terms of Causal Dynamical Triangulations (CDT), leading to the remarkable result
 that the large-scale curvature of the nonperturbatively generated quantum universe is compatible with that of a de Sitter space \cite{qrc3}.
On the one hand, this provides further evidence of the emergence of a universe with classical properties (in the sense of
expectation values) from a microscopic quantum dynamics.\footnote{Previous evidence is summarised in the reviews \cite{review1,review2}.} 
On the other hand, it demonstrates that the quantum Ricci curvature has good
averaging properties with respect to the microscopic quantum fluctuations that are present, producing a classical outcome (in the sense of
expectation values) on coarse-grained, macroscopic scales. 

In addition to providing a new test of the classical limit of quantum gravity, the quantum Ricci curvature gives us a new tool to characterize and quantify
the properties of quantum geometry in regimes far away from classicality. We do not know a priori what kind of nontrivial quantum behaviour the quantum 
Ricci curvature can exhibit, but there is no reason it should be any less complex than that of its classical counterpart. Until now, our know\-ledge of the quantum 
behaviour of the quantum Ricci curvature is limited. A natural testing ground are toy models of quantum gravity in two dimensions. Models of pure gravity,
whose action consists of a (topological) Einstein-Hilbert term and a cosmological-constant term, fall into two universality classes, distinguishable by the
Euclidean and Lorentzian nature of their metric signature. Both can be obtained as scaling limits of statistical models of random geometry, using
Dynamical Triangulations (DT) \cite{david,kkm} and Causal Dynamical Triangulations \cite{CDT0} respectively. Since gravity in two dimensions is trivial,
these models do not possess a nontrivial classical limit and therefore are maximally ``quantum". The averaged quantum Ricci curvature of
two-dimensional DT quantum gravity on a two-sphere was investigated in \cite{qrc2}, with the conclusion that it can be matched 
with good accuracy to that of a five-dimensional continuum sphere of constant curvature.\footnote{This is possible because the 
quantum Ricci curvature in two dimensions, which
contains the same information as the quantum Ricci scalar, is not constrained by the Gauss-Bonnet theorem \cite{qrc2}.} 
It is not straightforward to interpret this rather intriguing result in the absence of a classical limit and without an analytical derivation currently
at our disposal. Besides, since the quantum fluctuations are large, it is difficult to understand to what extent
the underlying global, spherical topology influences the quasi-local measurements of the quantum Ricci curvature. 

In this paper, we will investigate the quantum Ricci curvature of two-dimensional CDT quantum gravity on a torus, describing (1+1)-dimensional
universes whose spatial slices are compact one-spheres or circles of variable length $L$, 
and where for the convenience of the computer simulations, we cyclically identify the time direction. 
This model was solved analytically in \cite{CDT0} and is well studied; it has  
a spectral dimension $d_S$ of at most 2 and a Hausdorff dimension $d_H$ of almost surely 2 \cite{spec2d}, the latter in agreement with earlier theoretical \cite{CDT0,lessons} and numerical \cite{2dmatter} results. However, as already emphasized in \cite{2dmatter}, the fact that such dimensions have their
``classically expected" values,
equal to the topological dimension of the microscopic, triangular building blocks, does not imply that the geometry is locally flat. Plotting the
spatial volume of the universe as a function of (discrete) proper time for a typical member of the ensemble of geometries, one finds characteristic, strongly fluctuating ``candlestick" profiles (Fig.\ \ref{fig:candlestick}, see also Fig.\ \ref{fig:volume-profiles} below). This behaviour is not compatible with local flatness, 
which for the triangulated spacetimes of CDT would imply coordination 
number 6 for all vertices. However, it could still be the case that averaging the curvature over sufficiently large neighbourhoods yields an ``effectively flat" universe
on a larger scale. 

The new quantum Ricci curvature, which depends on a coarse-graining scale $\delta$ (roughly, the diameter of the averaging region), allows
us to establish whether or not such a flat behaviour is present, and at what scale. This is the issue we will investigate below. By virtue of the torus
topology, we will be able to distinguish between the (quasi-)local behaviour of the curvature, where only the geometry inside a contractible neighbourhood contributes,
and the global behaviour, which can involve some ``wrapping around" one or both of the compact torus directions. Lastly, the directional nature of the 
quantum Ricci curvature will allow us to explore the time- and spacelike directions of the quantum geometry separately, as was done previously in CDT
quantum gravity in four dimensions \cite{qrc3}. However, as already stressed above, unlike in four dimensions, there is no classical limit for quantum gravity in two dimensions, whose pure quantum character prevents us from formulating any straightforward expectation for its curvature properties.

\begin{figure}[t]
\vspace{-1cm}
\centerline{\scalebox{0.5}{\rotatebox{90}{\includegraphics[trim=2.5cm 1cm 2.5cm 1cm, width=0.95\textwidth, clip]{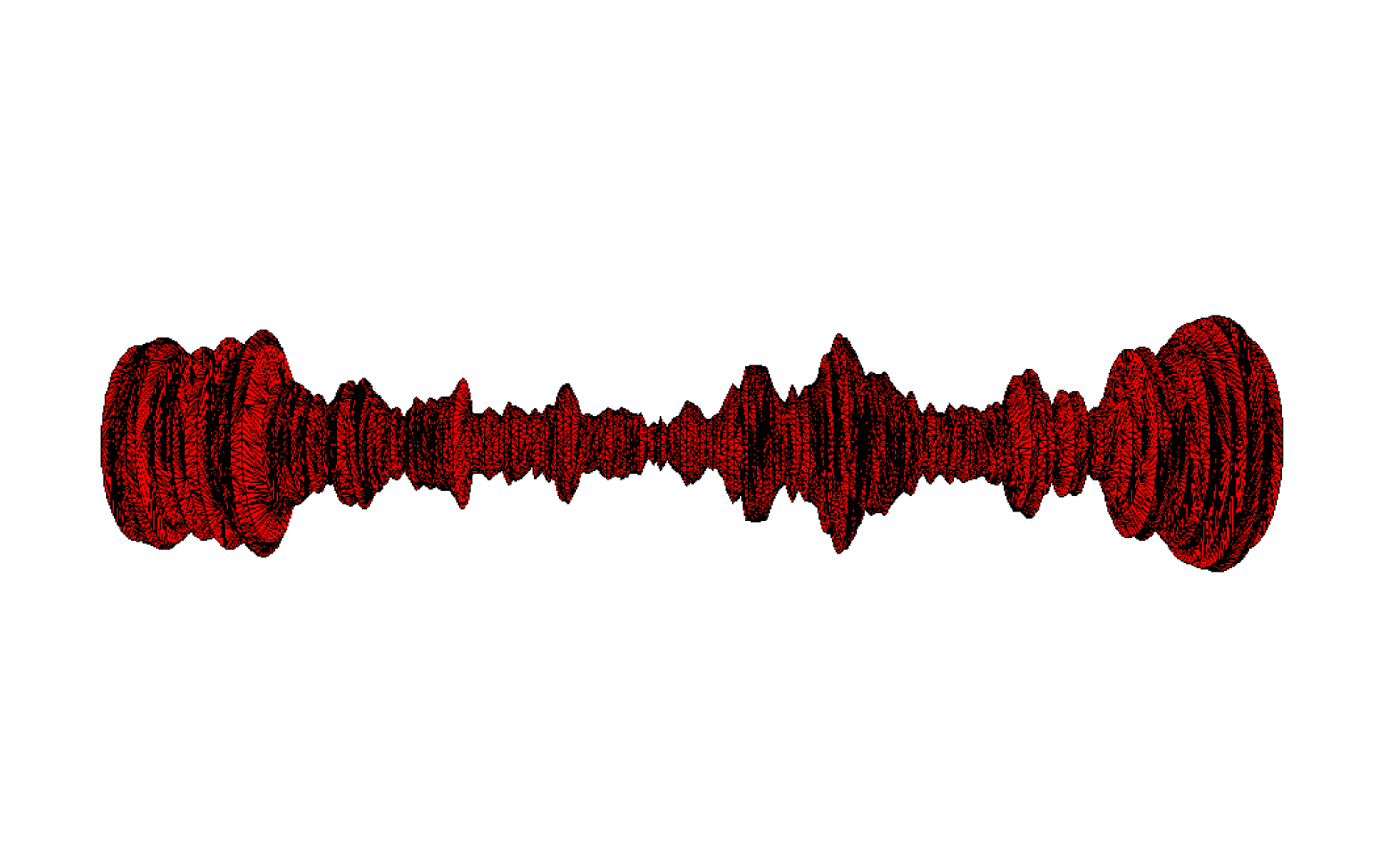}}}}
\vspace{-1cm}
\caption{Typical configuration contributing to the CDT path integral in two dimensions \cite{2dmatter}, 
with spacetime volume $N_2\! =\! 9408$, illustrating the fluctuating volume of 
the spatial $S^1$-slices as a function of proper time (horizontal direction, cyclically identified in simulations).}
\label{fig:candlestick}
\end{figure}

The structure of the remainder of this paper is as follows. In Sec.\ \ref{sec:2dcdt} we recall some relevant features and ingredients of Causal Dynamical
Triangulations in two spacetime dimensions.
Sec.\ \ref{sec:qrc} contains the definition and construction of the quantum 
Ricci curvature and the associated notion of the {\it curvature profile} \cite{qrc4}, as well as their use in the quantum theory. 
In Sec.\ \ref{sec:numerical}, we describe the set-up of the Monte Carlo simulations and present the curvature measurements.
To help us interpret the outcome, we perform a closer analysis of the influence of the torus topology
in Sec.\ \ref{sec:topology}, by determining quantitatively the fraction of data points that is affected by it.
This allows us to separate local and global aspects of the curvature profile and re-evaluate the previous measurements.
Some technical details have been relegated to appendix \ref{sec:appendix}.
Finally, Sec.\ \ref{sec:summary} contains a summary and conclusions.

\section{CDT quantum gravity in two dimensions}
\label{sec:2dcdt}

CDT quantum gravity was conceived as a nonperturbative implementation of the Lorentzian gravitational path integral, which involves a
sum over all
four-dimensional spacetime geometries with a physical, Lorentzian signature. It constituted a major advance over previous nonperturbative lattice
formulations\footnote{see \cite{livrev} for an overall assessment}, which for technical reasons addressed a different problem, 
namely, how to construct a quantum theory 
of gravity for four-dimensional Riemannian metric spaces, which lack a causal structure and with it the distinction between time-, space- and light-like
directions. It is important to realize that beyond perturbation theory on a flat background, where a Wick rotation is available, Lorentzian and Euclidean
quantum gravity are a priori different theories. To the extent they can be shown to exist as fundamental theories, there are no 
compelling arguments for why they should be simply related or even equivalent. 
If for some reason one wants to take unphysical, Riemannian metrics as a starting point, one must deal with the additional
difficulty of showing how Lorentzian concepts like time and causality can be retrieved from them. At this stage, the question is largely academic,
since nonperturbative Euclidean path integrals in four dimensions seem to suffer from generic pathologies and one has not found
evidence of second-order phase transitions, a prerequisite for the existence of a continuum limit of the underlying
regularized models. By contrast, four-dimensional CDT quantum gravity does possess higher-order phase transitions, and the investigation of
several quantum observables (dynamical dimensions, volume and curvature profiles) in CDT has revealed highly nontrivial evidence for the emergence 
of a classical limit from its Planckian quantum dynamics (see \cite{review2} for a recent review).  

Since two-dimensional quantum gravity is merely a toy model for the real theory, there is no compelling physical reason to study it in any particular signature.
However, a pivotal aspect of the analytic solution of CDT in two dimensions \cite{CDT0} was to provide a first explicit example of the inequivalence between Lorentzian 
and Euclidean gravitational path integrals, which manifests itself in different values for several universal critical exponents, including fractal dimensions and
the string susceptibility \cite{lessons,nato2d}. Following the original, purely geometric CDT model, there have been numerous analytical and numerical studies of
coupled Lorentzian gravity-matter systems in two spacetime dimensions, 
involving spin systems \cite{2dmatter,crossing,shaken}, dimers \cite{dimercdt1,dimercdt2}, scalar fields 
\cite{scalars,specscalar}, gauge fields \cite{gauge2d,candido} and loop models \cite{loopmodel} 
(see these papers for more extensive bibliographies). Other work in two dimensions has dealt with
the relation between the Euclidean and Lorentzian quantum geometries \cite{relation}, a possible interpolation between both models 
involving ``locally causal dynamical triangulations" \cite{lcdt}, the universal nature of CDT quantum gravity \cite{universality},
and a generalized version of CDT, where spatial topology changes are permitted \cite{cap,sft}. 

\begin{figure}[t]
	\centering
	\includegraphics[width=0.9\textwidth]{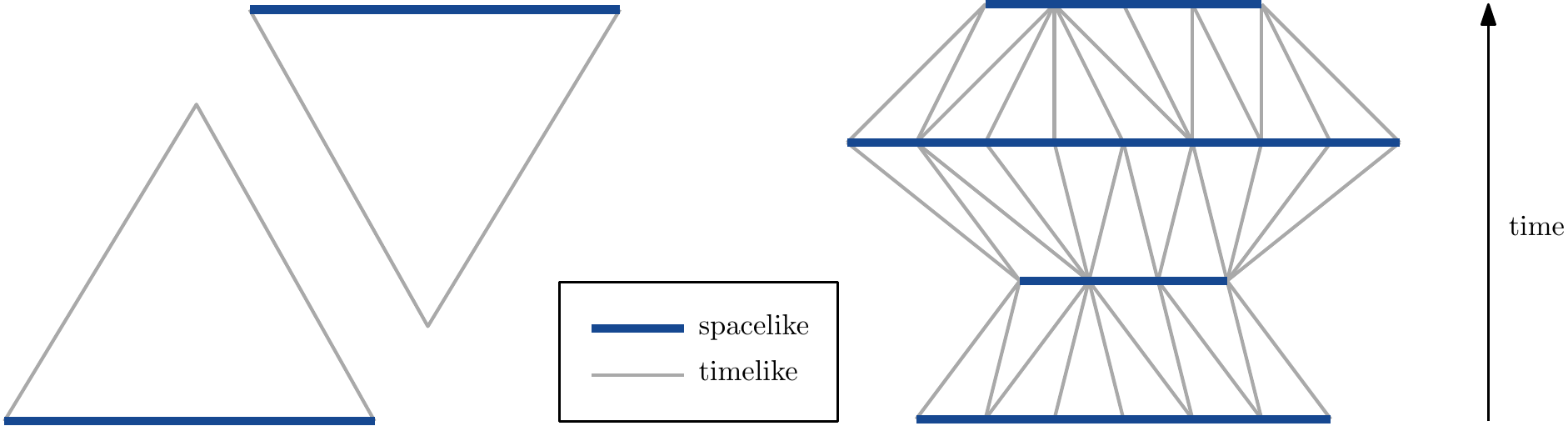}
	\caption{Minkowskian isosceles triangles with one space- and two timelike edges (left) are the elementary building blocks of the 
	piecewise flat lattices of CDT. They are assembled into curved, triangulated spacetimes with a stacked structure (right). Note that the
	graphical representation is not isometric.}
	\label{fig:cdt-sample}
\end{figure}

To set the stage for our investigation of the quantum Ricci curvature, we will briefly recap the ingredients of CDT in two dimensions (see \cite{npb,review1}
for more comprehensive technical details). The 
gravitational path integral in this approach is defined as the continuum limit of a sum over regularized, curved spacetimes, given in terms of 
simplicial manifolds, the {\it causal triangulations}. In order to evaluate this nonperturbative path integral, either analytically or numerically, one needs
to analytically continue it, using the Wick rotation that is a key feature of this formulation \cite{CDT0,review2}. The CDT path integral after the Wick rotation
has the form of a real-valued partition function
\begin{equation}
	Z = \sum_{T} \frac{1}{C_T} \, {\rm e}^{-S_\lambda[T]}, \quad \quad S_\lambda[T] = \lambda\, N_2(T),
	\label{eq:partition-sum}
\end{equation}
where the sum is over causal triangulations $T$ of topology $S^1\times S^1$ and $C_T$ denotes the order of the automorphism group of $T$.
Since the topology is fixed, the Euclidean gravitational action $S_\lambda [T]$ consists only of a cosmological-constant term, 
where the bare cosmological constant $\lambda$ multiplies
the volume $N_2(T)$, counting the number of triangles contained in the triangulations $T$. The simple form of $S_\lambda$ comes about because
all triangulations are built from a single type of triangular building block of fixed geometry, a flat Minkowskian isosceles triangle with one spacelike
and two timelike edges, which in the causal triangulations can appear with two different time orientations (Fig.\ \ref{fig:cdt-sample}, left). The Wick 
rotation of CDT turns
this into a flat Euclidean triangle, which without loss of generality can be taken to be equilateral. Note that the purely Euclidean path integral in terms of
DT, whose continuum limit is Liouville gravity, also uses flat equilateral triangles and formally looks identical to eq.\ (\ref{eq:partition-sum}), but is defined
on a different configuration space of simplicial manifolds, which do not carry any imprint of the causal structure of CDT. The latter is captured
by causal ``gluing rules" for the elementary building blocks, which result in a stacked structure, a discrete analogue of global hyperbolicity 
(Fig.\ \ref{fig:cdt-sample}, right).\footnote{Note that the ``straightness" of the one-dimensional simplicial submanifolds of the spacelike edges 
(thick lines) in Fig.\ \ref{fig:cdt-sample} is a feature of the graphic representation and does not indicate the absence of extrinsic curvature of
these submanifolds (see \cite{review2} for a detailed discussion of the roles of time and causality in CDT).}

Despite being sometimes labelled a ``discrete approach", CDT quantum gravity is not based on a postulate of fundamental discreteness at the
Planck scale, but uses simplicial building blocks merely as part of a lattice-like regularization of the quantum theory.   
The edge length $a$ of such a simplex or triangle provides a short-distance (UV) cutoff, which in any continuum limit will be sent to zero, accompanied by
a suitable renormalization of the bare coupling constant(s). 
In numerical simulations in two dimensions, the limit $a\rightarrow 0$ is realized by taking $N_2\rightarrow\infty$ for the total number of triangles,
where $a\propto 1/\sqrt{N_2}$. Since the limit of infinite system size cannot be attained with finite computational resources, the behaviour of 
an observable in the continuum limit is extrapolated by studying it on
a sequence of systems of large and increasing $N_2$ using finite-size scaling, a standard tool in the analysis of statistical
systems \cite{MC}. In Monte Carlo studies of any physical quantities, including the quantum Ricci curvature considered below, 
data points at or very near to the cutoff scale are usually discarded, because they are affected by the details of the chosen 
regularization, so-called lattice artefacts. Instead, one is interested in universal, regularization-independent
properties that manifest themselves on scales $\gg a$, and which are characterised by a specific scaling behaviour as a
function of the system size.

\section{Quantum Ricci Curvature}
\label{sec:qrc}

The introduction of the quantum Ricci curvature \cite{qrc1} was motivated by the quest for a notion of curvature applicable to
quantum gravity in a nonperturbative regime. Its geometric implementation is based on the observation that the Ricci curvature of a smooth Riemannian
space can be extracted by comparing the distance between two nearby infinitesimal spheres with the geodesic distance of their
centres. In a positively curved space, the average geodesic distance between pairs of points from the two spheres (related to 
each other pairwise by parallel transport \cite{ollivier1}) is
smaller than the distance between the two sphere centres, and the converse is true in a negatively curved space. 
This insight was used by Ollivier to define a generalized, ``coarse-grained" notion of Ricci curvature, applicable to a much larger class of metric
spaces, including nonsmooth and discrete ones \cite{ollivier}. 

\begin{figure}[tb]
	\centering
	\includegraphics[width=0.5\textwidth]{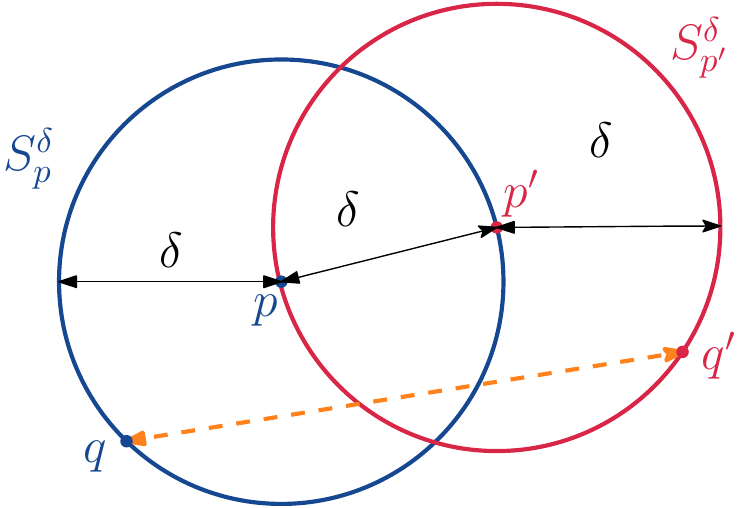}
	\caption{The double-circle configuration involved in computing the average sphere distance $\dbar(S_p^\delta, S_{p'}^\delta)$ of eq.\ (\ref{sdist}) 
	on a two-dimensional Riemannian manifold.
	}
	\label{fig:avg-sphere-dist}
\end{figure}

To apply this idea in the context of quantum gravity and CDT in particular, additional considerations must be taken into account. The requirement to go beyond the
cutoff scale to obtain a meaningful notion of (quasi-)local quantum Ricci curvature is implemented by considering pairs of
spheres whose radius in lattice units is given by an integer $\delta$ much larger than 1. Since the Wasserstein distance -- based on the idea of
optimal transport -- used in the construction of the Ollivier-Ricci curvature \cite{ollivier} becomes computationally very expensive 
with increasing $\delta$, the construction of the quantum Ricci curvature uses a different notion of sphere distance, the
{\it average sphere distance}.\footnote{Our philosophy here is different from that of ``combinatorial quantum gravity" \cite{tru2,tru3,tru4,tru5},
which investigates the emergence of geometry from specific ensembles of random graphs, 
with the ambition of having its Lorentzian structure emerge alongside, something not yet 
achieved by any model as far as we know. In this approach, one uses the
Ollivier-Ricci curvature in its original form for small values of $\delta$, and has recently also compared it to another discrete notion of
so-called Forman-Ricci curvature \cite{forman}.} On a $D$-dimensional Riemannian space with 
metric $g_{\mu\nu}(x)$, and for a pair $S_p^\delta$, $S_{p'}^\delta$ of
spheres (circles for $D\! =\! 2$) of radius $\delta$, whose centres $p$ and $p'$ are a distance $\delta$ apart\footnote{These are the choices made for
the quantum Ricci curvature \cite{qrc1}. In principle one could choose the centre distance different from the sphere radii, but for the purposes
of the quantum theory, our primary aim is to construct a notion of curvature associated with a single (coarse-graining) scale $\delta$.}, 
the average sphere distance $\bar{d}(S_p^{\delta},S_{p'}^{\delta})$ is defined as
\begin{equation}
\bar{d}(S_p^{\delta},S_{p'}^{\delta}):=\frac{1}{\textit{vol}\,(S_p^{\delta})}\frac{1}{\textit{vol}\,(S_{p'}^{\delta})}
\int_{S_p^{\delta}}d^{D-1}q\; \sqrt{h} \int_{S_{p'}^{\delta}}d^{D-1}q'\; \sqrt{h'}\ d_g(q,q'),\;\;\;\;\; d_g(p,p')=\delta.
\label{sdist}
\end{equation}
In other words, one integrates the geodesic distance $d_g(q,q')$ over all point pairs $(q,q')\in S_p^\delta \times S_{p'}^\delta$ from
the two spheres, and normalizes the resulting double integral by the $(D-1)$-dimensional sphere volumes $vol(S)$, computed with the help of the induced metrics $h$
on the spheres (see Fig.\ \ref{fig:avg-sphere-dist} for a schematic illustration of the two-dimensional case, where the dashed orange line indicates
the shortest path between the points $q$ and $q'$).
In the Riemannian case, this prescription for computing the distance between the two spheres leads to results equivalent to those using the Wasserstein distance, 
in the sense that one can recover the 
standard local Ricci curvature from an infinitesimal expansion \cite{qrc2}, just like for the Ollivier-Ricci curvature, and that for non-infinitesimal $\delta$ the 
behaviour of the two sphere distances is qualitatively similar, at least on constantly curved classical spaces \cite{qrc1}.

\begin{figure}[ht]
	\centering
	\includegraphics[width=0.6\textwidth]{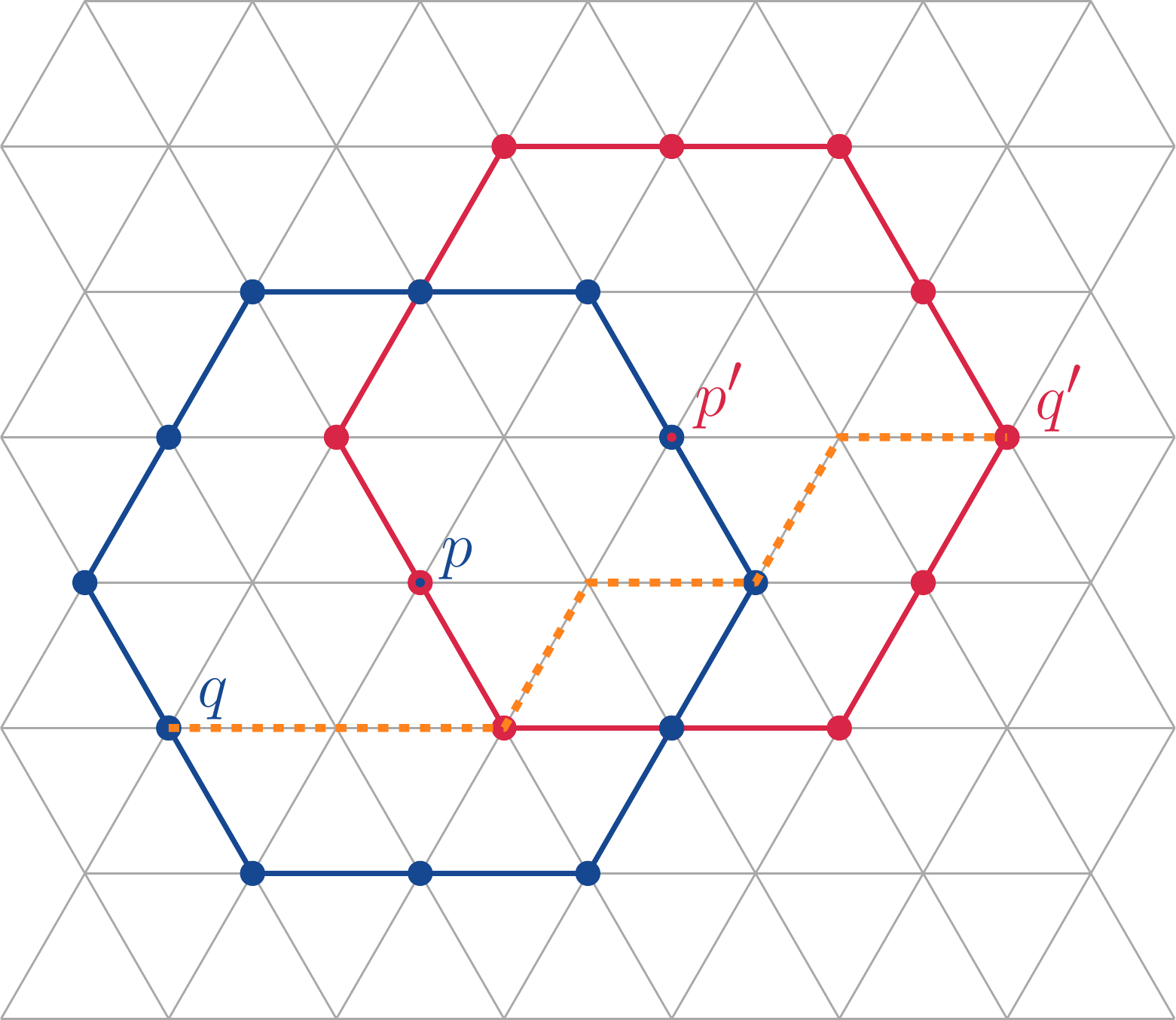}
	\caption{The double-circle configuration involved in computing the average sphere distance $\dbar(S_p^2, S_{p'}^2)$ of eq.\ (\ref{eq:disc-avg-sph-dst}) 
	on the two-dimensional regular triangulation of the flat plane, using the link distance.}
	\label{fig:cdt-sphere-pair}
\end{figure}

On a two-dimensional piecewise flat manifold $T$ in the CDT ensemble, one can implement formula (\ref{sdist}) in a straightforward way, using discrete analogues of 
distances and volumes, yielding
\begin{equation}
	\dbar (S^\delta_{p},S^\delta_{p'} ) = \frac{1}{N_1(S^\delta_{p})} \frac{1}{N_1(S^\delta_{p'})} \sum_{q \in S^\delta_{p}} \sum_{q' \in S^\delta_{p'}} d(q,q'), 
	\;\;\; d(p,p')=\delta,
	\label{eq:disc-avg-sph-dst}
\end{equation}
where we continue to use the notation $\dbar$ for the average sphere distance, and $q$, $q'$, $p$, $p'$ now denote vertices of $T$.
The standard notion of distance on an equilateral triangulation $T$ is the integer-valued link distance $d(q,q')$ between pairs of vertices $q$, $q'$ in $T$,
counting the number of links (edges) in the shortest path of contiguous links joining $q$ and $q'$. By definition, a ``sphere" $S_p^\delta$ centred at the
vertex $p\in T$ is the set of all vertices $q$ at link distance $\delta$ from $p$, and $N_1(S_p^\delta)$ counts the number of vertices
in this set. These sets only loosely resemble the round spheres of the smooth continuum, as is illustrated by Fig.\ \ref{fig:cdt-sphere-pair},
which depicts a pair of spheres on a regular tessellation of the plane in terms of equilateral triangles, forming a hexagonal lattice. 
The dashed orange line indicates a shortest
path between the vertices $q$ and $q'$. As happens generically, this path is not unique. Note that according to our definition, the spheres
consist only of the vertices, and not the links between neighbouring vertices, which in Fig.\ \ref{fig:cdt-sphere-pair} 
have merely been included to guide the eye. 
On generic CDT configurations, which are much less regular, one cannot in general link vertices of a ``sphere" pairwise in a way that
results in a single loop without intersections or overlaps.

Later we will occasionally work with the dual link distance, which is defined on the dual vertices of the triangulation (the centres of the triangles), 
and counts the number of dual links in the shortest path between a pair of dual vertices. The dual links can be thought of as the edges of the trivalent 
graph dual to a given triangulation $T$. By virtue of universality, the link distance and the dual link distance should lead to the same notion of physical distance
in the continuum limit, up to an overall scale due to the difference between the discrete link and dual link units.\footnote{We are not aware of counterexamples
in either DT or CDT models of quantum gravity in any dimension.} One would therefore also expect equivalent physical
results for the average sphere distance (and the quantum Ricci curvature extracted from it) in the continuum limit, independent of the choice of
the discrete distance function. However, since the actual simulations always take place on finite lattices, it can be convenient to make a specific choice.
For example, in the recent investigation of the quantum Ricci curvature in CDT quantum gravity in four dimensions, in order to obtain measurements
with a sufficiently fine distance resolution at the given lattice size, it turned out to be important to use the dual link distance \cite{qrc3}.    

Only a small step is required to obtain the quantum Ricci curvature\footnote{The ``$q$" in $K_q$ stands for ``quantum" and does not denote a
point or vertex.} $K_q(p,p')$ from the average sphere distance $\bar{d}(S_p^{\delta},S_{p'}^{\delta})$
associated with a point pair $(p,p')$ at distance $\delta$. 
Note that such a point pair can be thought of as a generalization of the notion of a tangent vector at the point $p$, which
``points from $p$ to $p'$" and reflects the directional dependence of the quantum Ricci curvature. To obtain the
quantum Ricci curvature, we take the quotient of two distances, namely, the average sphere distance and the distance $\delta$,  
\begin{equation}
\bar{d}(S_p^{\delta},S_{p'}^{\delta})/\delta=c_q\, (1 - K_q(p,p')).
\label{qric}
\end{equation}
Considering this expression as a function of $\delta$ (while keeping the ``direction" $\overline{pp'}$ fixed), the factor $c_q$ by definition
describes its constant, $\delta$-independent part, while $K_q(p,p')$ captures any nontrivial $\delta$-dependence. In the
Riemannian case, one has $\overline{pp'}\! =\!\delta v$ for some unit tangent vector $v$ at $p$ and sufficiently small $\delta$,
and $c_q$ is given by the limit $c_q\! :=\!\lim_{\delta\rightarrow 0} \bar{d}/\delta$. In this case, $c_q$ is a constant that only depends on 
the dimension $D$ of the manifold, with $c_q\! =\! 1.5,\, 1.5746,\, 1.6250,\, 1.6524$ in dimension $D\! =\! 1,\, 2,\, 3,\, 4$ respectively,
and $K_q$ is to lowest order proportional to $\delta^2$ \cite{qrc2}. For example, in $D\! =\! 2$ one finds
\begin{equation}
\bar{d}/\delta= 1.5746+\delta^2 \left(-0.1440\, \textit{Ric}(v,v)+{\cal O}(\delta)\right),
\label{2dexp}
\end{equation}
where $\mathit{Ric}(v,v)\! =\! R_{ij}v^iv^j$ is the usual Ricci curvature, evaluated on the unit vector $v$ at the point $p$ 
in the direction of $p'$. Note that in two dimensions $\mathit{Ric}(v,v)$ coincides with the Gaussian curvature at $p$. 

On triangulations, the (dual) link distance $\delta$ is integer-valued and the limit $\delta\!\rightarrow\! 0$ is not defined. Moreover, as we have
already pointed out, measurements of the average sphere distance at small values of $\delta$ are afflicted by lattice artefacts.
One can then define $c_q\! :=\! (\bar{d}/\delta)|_{\delta=\delta_0}$, where $\delta_0$ is the distance above which lattice artefacts are
negligible. Numerous investigations of the average sphere distance on triangulated and other piecewise flat manifolds (regular lattices
and Delaunay triangulations
\cite{qrc1}, DT \cite{qrc2} and CDT \cite{qrc3} configurations) have found that $\delta_0\approx 5$ in terms of the link distance, with
larger values of $\delta_0$ when the dual link distance is used. These results also show that the value of $c_q$ is not universal, but
depends on the details of the lattice discretization. This includes the lattice direction in which the quantum Ricci curvature is evaluated, 
and is related to the absence of exact rotational
invariance around a given lattice vertex. It is similar to the ``staircase effect" on a regular tiling of the two-dimensional flat plane
by identical squares. In this case, the link distance is anisotropic, since walking $\delta$ steps from a vertex $p$ along a shortest path in 
the diagonal direction results in a zigzag path that only covers a distance $\delta/\sqrt{2}$ in the underlying flat plane (assuming the link length is 1), 
while shortest paths of $\delta$ steps in the horizontal or vertical direction cover a continuum distance of $\delta$ \cite{qrc3,thesisklitgaard}. 

Exploring the quantum Ricci curvature $K_q$ in the full quantum theory, which is based on a nonperturbative path integral over all spacetimes, requires that we 
construct diffeomorphism-invariant observables $\cal O$ depending on $K_q$. We can then try to determine their expectation
values
\begin{equation}
\langle {\cal O}\rangle =\frac{1}{Z}\,\sum_T \frac{1}{C_T}\, {\cal O}[T]\, {\rm e}^{-S_\lambda [T]}
\label{expect}
\end{equation}
with the help of Monte Carlo simulations, where the partition function $Z$ was defined in eq.\ (\ref{eq:partition-sum}). 
As explained in detail elsewhere \cite{qrc4,review2}, invariant observables in pure gravity generically involve a spacetime averaging, to
eliminate the dependence on individual points (or other subregions), which are identified by their unphysical coordinate labels. 
A suitable classical, global curvature observable of this kind, the {\it curvature profile}, was introduced in \cite{qrc4}. To eliminate the 
dependence of the quasi-local quantum Ricci curvature $K_q(p,p')$ on its local arguments, the locations of the circle centres $p$ and $p'$ in the
average sphere distance are integrated (or in the discrete case summed) over, subject to the constraint that the distance
between $p$ and $p'$ is always equal to $\delta$. 
The resulting normalized average of the average sphere distance at the scale $\delta$ is given by
\begin{equation}
\bar{d}_{\rm av}  (\delta):=\frac{1}{{\cal N}_\delta}
\int_{M}d^2p\, \sqrt{g} \int_M d^2 p' \sqrt{g}\;\,  \bar{d}(S_p^{\delta},S_{p'}^{\delta})\; \delta_D(d_g(p,p'),\delta),
\label{avdist}
\end{equation}   
where $\delta_D$ denotes the Dirac delta function and we have adopted a continuum language appropriate in the presence of a 
Riemannian metric $g_{\mu\nu}$ on a two-dimensional manifold $M$. 
On a triangulated manifold $T$, the integrals must be substituted by sums over vertices in $T$, the
geodesic distance $d_g$ by the link distance or dual link distance $d$, and the Dirac delta function by a discrete Kronecker delta $\delta_K$, 
leading to\footnote{By a minor abuse of notation, we use the same symbol $\bar{d}_{\rm av}$ for both the smooth and the piecewise
flat case. This should not lead to any confusion, since only the latter will be relevant from Sec.\ \ref{sec:numerical} onwards.} 
\begin{equation}
\bar{d}_{\rm av}  (\delta):	=  \frac{1}{{\cal N}_\delta} \sum_{p \in T}\sum_{p' \in T} \dbar (S_p^\delta, S_{p'}^\delta) \, \delta_K(d(p,p'), \delta).
	\label{avdisttri}
\end{equation}
The normalization factor ${\cal N}_\delta$ is given by
\begin{equation}
{\cal N}_\delta=\int_{M}d^2x \sqrt{g} \int_M d^2 x' \sqrt{g}\;\; \delta_D(d_g(x,x'),\delta)
\label{zdelta}
\end{equation}
in the smooth case, eq.\ (\ref{avdist}), and by the double sum 
\begin{equation}
	{\cal N}_\delta = \sum_{p \in T}\sum_{p' \in T}  \delta_K(d(p,p'), \delta)
\end{equation}
in the piecewise flat case, eq.\ (\ref{avdisttri}).
The expressions (\ref{avdist}) and (\ref{avdisttri}) include averages over directions, and therefore
allow us to extract an (averaged) \textit{quantum Ricci scalar} $K_{\rm av}(\delta)$ 
from the curvature profile, which is defined by the quotient
\begin{equation}
\bar{d}_{\rm av}(\delta)/\delta=:c_{\rm av} (1 - K_{\rm av}(\delta)).
\label{profile}
\end{equation}
In the continuum case, the constant $c_{\rm av}$ is defined by $c_{\rm av}\! :=\!\lim_{\delta\rightarrow 0} \bar{d}_{\rm av}/\delta$,
with a suitable generalization for piecewise flat triangulations. If an intrinsic notion of direction is present, it may not be necessary to
perform an averaging over all directions to obtain a curvature observable.
CDT configurations carry a piecewise flat Lorentzian
structure that allows us to distinguish between time- and spacelike directions, even after the Wick rotation. 
It means that one can capture
direction-dependent information about the Ricci curvature by summing only over a subset of centre pairs $(p,p')$, which satisfy
specific restrictions with regard to their space- or timelike separation, similar to what was done in \cite{qrc3}.   

Finally, in the quantum theory we will be interested in the expectation value of the 
nonlocal curvature profile ${\cal O}\! =\! \bar{d}_{\rm av}(\delta)/\delta$ (and the quantum Ricci curvature
extracted from it) as a function of $\delta$, obtained by taking the ensemble average over CDT geometries $T$,
\begin{equation}
	\langle \bar{d}_{\rm av}(\delta)/\delta \rangle \equiv \langle \bar{d}_{\rm av}(\delta)\rangle /\delta 	
	= \frac{1}{\delta}\, \sum_{T } \frac{1}{C_T}\,  \bar{d}_{\rm av}(\delta)\, e^{- S_\lambda [T]}   .
	\label{eq:curv-prof-cdt}
\end{equation}

\section{Measuring curvature profiles}
\label{sec:numerical}

Although the partition function \eqref{eq:partition-sum} of two-dimensional CDT quantum gravity can be evaluated analytically \cite{CDT0}, 
it is not known at this stage how to compute the expectation value (\ref{eq:curv-prof-cdt}) of the curvature profile exactly. 
To understand its quantitative behaviour, we therefore turn to computer simulations. As outlined above, we will measure this observable
at various fixed spacetime volumes, in our case in the range $N_2\in [50k,600k]$, and try to extract its continuum behaviour from a finite-size scaling analysis.  

Our procedure used a Markov chain Monte Carlo algorithm (our implementation code can be found at \cite{2dcdtgithub}) to generate a sequence of independent CDT configurations $\{ T_n\}$, $n\! =\! 1,2,\dots ,N$, all 
with torus topology $T^2$
and fixed numbers $N_2$ of triangles and $\tau$ of spatial slices. This sequence serves as a random sample of the probability distribution given 
by the partition function \eqref{eq:partition-sum} at fixed volume. 
Since the computation of the exact sphere distance $\bar{d}_{\rm av}  (\delta)$ of eq.\ (\ref{avdisttri}) for a given triangulation $T_n$ is very 
costly\footnote{A separate BFS has to be performed for every vertex in $S_p^\delta$, with a maximum depth of $3\delta$,
the upper bound for the distance between any pair of points $(q,q')\! \in\! S_p^\delta\! \times\! S_{p'}^\delta$, corresponding to a path through $p$ and $p'$
(c.f.\ Fig.\ \ref{fig:avg-sphere-dist}). 
A simple but effective optimization of the BFS is to relabel the spheres such that the ``first" sphere $S_p^\delta$ is the one containing fewer vertices.}, 
we only compute a sample of sphere distances $\dbar(S_p^\delta, S_{p'}^\delta)$ on each $T_n$, one for each $\delta$ in the chosen 
integer range $\delta\!\in\! [1,\delta_{\rm max}]$, which will then contribute to the ensemble average (\ref{eq:curv-prof-cdt}) at the given $\delta$-value.

For given $T_n$ and $\delta$, a pair $(S_p^\delta, S_{p'}^\delta)$ of spheres is constructed by first picking a vertex $p\!\in\! T_n$ with uniform probability, 
and finding the set $S_p^\delta$ of all vertices at link distance $\delta$ from the centre $p$ by a breadth-first search (BFS) algorithm. 
One then picks uniformly at random a vertex $p'\! \in\! S_p^\delta$, and constructs the set $S_{p'}^\delta$ in the same manner. 
Given such a pair of spheres, the average sphere distance $\dbar(S_p^\delta, S_{p'}^\delta)$ is determined exactly by carrying out a BFS starting from every point 
$q \in S_p^\delta$ until the distances between all pairs $(q,q') \in S_p^\delta \times S_{p'}^\delta$ have been found. 
Repeating this procedure for each triangulation in the sequence $\{ T_n\}$, we obtain $N$ average sphere distance measurements for every 
value of $\delta$, whose average yields a numerical estimate of the expectation value $\langle \bar{d}_{\rm av}(\delta)\rangle$ and thus of the
curvature profile.

As part of the same Monte Carlo simulation, we also collected average sphere distances using the dual link distance. For each configuration $T_n$
of some sequence $\{ T_n\}$, $n\! =\! 1,2,\dots ,N$, and for each dual distance $\delta\in [1,\delta^*_{\rm max}]$, 
we randomly picked a dual vertex $p^*$ (unrelated to any vertex $p$
selected for the link-distance measurements), and subsequently a dual vertex $p'^*$ from the $\delta$-sphere $S_{p^*}^\delta$ around $p^*$, etc.,
following a procedure completely analogous to the one just described to extract the average sphere distance $\dbar(S_{p^*}^\delta, S_{p'^*}^\delta)$
with respect to the dual link distance. Since the dual graph is trivalent, the size of the ``dual" spheres $S_{p^*}^\delta$, based at a dual vertex $p^*$,
grows on average more slowly as a function of $\delta$ than the size of the spheres $S_p^\delta$ based at a vertex $p$. By the same token, one traverses faster
through a given triangulation $T$ along a geodesic when taking steps along links rather than along dual links, a well-known phenomenon that leads to a relative scaling
between the two notions of lattice distance\footnote{roughly speaking, a factor of order 2}, compared with any particular notion of (renormalized) geodesic 
distance in the continuum limit. We therefore expect that the $\delta$-range for which we can determine the curvature profile of triangulations of
a given size $N_2$ is larger when we use the dual link distance, in other words, that $\delta^*_{\rm max}>\delta_{\rm max}$, which indeed turns out to be the case.

A prominent feature of the quantum Ricci curvature is its dependence on a length scale $\delta$, captured by the notion of the curvature profile. 
In the context of the nonperturbative CDT quantum theory, and depending on the (finite) lattice size and resolution, this scale allows us in principle to 
distinguish various
regimes: an unphysical regime below the cutoff, a Planckian regime for small $\delta$ above the cutoff, and a (semi-)classical regime for larger $\delta$.
In the region where $\delta$ becomes comparable to the linear system size, the quantum Ricci curvature will by construction ``feel" the global structure of the 
underlying spacetime, which is determined by our choice of boundary conditions. 

Recall that in nongravitational lattice field theories, one often works with cubic $D$-dimensional lattices, where all directions are identified periodically, yielding a $D$-dimensional torus $T^D$ topologically. If one wants to approximate physics in Minkowski space
(after Wick rotation, in flat Euclidean space), such a compactification does not reflect a physical property, but is chosen for simplicity, and forced upon 
us by the finite size of the system (unless one wants to deal with open boundaries).
When measuring observables, one therefore tries to minimise any dependence on this unphysical global structure. 

The situation in nonperturbative lattice gravity is slightly different. Firstly, since gravity is dynamical, the geometric properties of the lattice are
not fixed a priori. Even if the total spacetime volume is kept fixed, a concept like the ``linear size" of the spacetime is not
fixed at the outset and at the very least subject to quantum fluctuations. Besides, geometric quantities like lengths, areas and volumes may not scale
``canonically" in the continuum limit, i.e. not as expected from the topological dimension of the discrete building blocks that were used to
construct the regularized theory. The classic example is two-dimensional Liouville gravity (Euclidean DT), whose quantum geometry is characterised 
by a Hausdorff dimension of four, and not two as one may have expected na\"ively \cite{dim2d,moredim2d,dim2dnum}. 
Secondly, since gravity is a theory of spacetime, one may be interested in studying specifically the influence of the global spatial or spacetime
topology on the gravitational dynamics, regarding it as a feature rather than a technical necessity. 

Returning to the discussion of the quantum Ricci curvature, its built-in scale dependence means that even in the classical case, it can be used to probe 
not only an infinitesimal, local neighbourhood, producing results compatible with a small-$\delta$ expansion like eq.\ (\ref{2dexp}), but also 
noninfinitesimal $\delta$, giving rise to an averaged or coarse-grained Ricci curvature associated with a macroscopic length scale $\delta$. 
For general metrics, this regime is difficult to access analytically, an issue that is at the heart of the so-called averaging problem in classical
cosmology (see e.g.\ \cite{average}). Addressing this problem in terms of the quantum Ricci curvature is of independent interest in the 
classical theory and an issue we will return to elsewhere. 

In the two-dimensional quantum theory we are studying presently, it is clear that
the results for the curvature profile will carry an imprint of the global toroidal topology whenever the sphere radius $\delta$ becomes sufficiently large. 
A feature of the torus we will exploit below is that -- unlike for the two-sphere -- one can formulate sharp criteria for when a particular average sphere distance 
measurement is affected by the nontrivial topology. We already mentioned that the linear extension of the ``double sphere" $(S_p^\delta, S_{p'}^\delta)$ 
is $3\delta$. If one wants to exclude global topological effects, one should clearly avoid that the area enclosed by the two spheres wraps around the 
torus and self-overlaps in a nontrivial way. However, this is not sufficient. One must also make sure that there are no geodesics between any pair of points
$(q,q')\! \in\! S_p^\delta \times S_{p'}^\delta$ that represent shortcuts by wrapping around the ``back" of the torus and that would not be present if the
torus was cut open appropriately outside the double sphere region (see Sec.\ \ref{sec:topology} for further discussion, as well as related considerations in \cite{qrc4}). 
{\it If} we were on a fixed, flat torus in the continuum,
obtained by identifying opposite sides of a flat rectangle with side lengths $l_1$ and $l_2$, a sufficient condition for this not to happen would be
$6\delta\!\leq\! l_i$. The same condition applies to CDT configurations with respect to the periodically identified time direction. Namely, if we want
to exclude any influence of the compactified time on the quantum Ricci curvature measurements, we should choose $\delta_{\rm max}$ such that
$6\delta_{\rm max}\! \leq\! \tau$, where $\tau$ is the (fixed) time extension of a CDT geometry, measured in terms of discrete proper time. 
When using the dual link distance in the construction of the quantum Ricci curvature, the corresponding condition is $3\delta^*_{\rm max}\!\leq \!\tau$.  
To determine when and how the compactness in the spatial direction on the torus influences the curvature measurements is much more involved, since the
volume of the spatial slices is subject to large quantum fluctuations. 
Results for the curvature profile that are independent of the boundary condition in the time direction will be presented in Sec.\ \ref{sub:results},
while the influence of the spatial compactness will be analyzed in detail in Sec.\ \ref{sec:topology}.

\subsection{Measurement results}
\label{sub:results}

Following the procedure outlined above, we have determined the expectation value $\langle \bar{d}_{\rm av}(\delta)/\delta \rangle$ 
of the curvature profile on toroidal CDT geometries for two different time extensions $\tau$ and a range of spacetime volumes $N_2$, as a function
of the link distance. 
For configurations with $\tau\! =\! 183$ spatial slices, measurements were performed in the range $\delta\! \in\! [1, \delta_{\rm max}\! =\! 30]$, 
to avoid any direct influence of the periodic identification in time, a boundary condition chosen merely for the convenience of
the simulations. 
We have collected $50k$ measurements at volumes $N_2\! =\! 50$, 70, 150 and 250$k$, and $30k$ measurements at $N_2\! =\! 350k$, where 
increasing the total volume for fixed $\tau$ results in a larger average size of the spatial slices at integer time.
In addition, we performed $30k$ measurements for configurations with time extension $\tau\! =\! 243$, 
using an extended $\delta$-range with $\delta_{\rm max}\! =\! 40$ and 
a volume of $N_2\! =\! 600k$. We also determined $\langle \bar{d}_{\rm av}(\delta)/\delta \rangle$ in terms of the dual link distance, 
for $\tau\! =\! 183$, $\delta^*_{\rm max}\! =\! 60$ and at volumes 
$N_2\! =\! 100$, 200 and $250k$, based on $50k$ data points. In between measurements, we performed $N_2\!\times\! 1000$ attempted moves. Considering that the equilibrium acceptance ratios are around $10\%$, this implies that every site of the geometry was updated roughly 100 times on average in between two sampled triangulations $T_n$ and $T_{n+1}$.

We used a data blocking procedure to check for autocorrelations in the data collected at subsequent measurement steps, but such correlations were found to be absent. This is perhaps not too surprising, given that a single average sphere distance measurement is quasi-local (as long as $\delta$ does not become too large
compared to the system size), and we have chosen to take only one
data point for every $\delta$ for a given configuration $T_n$, each one from a double-sphere with different initial point $p$.

\begin{figure}[t]
	\centering
	\includegraphics[width=0.7\textwidth]{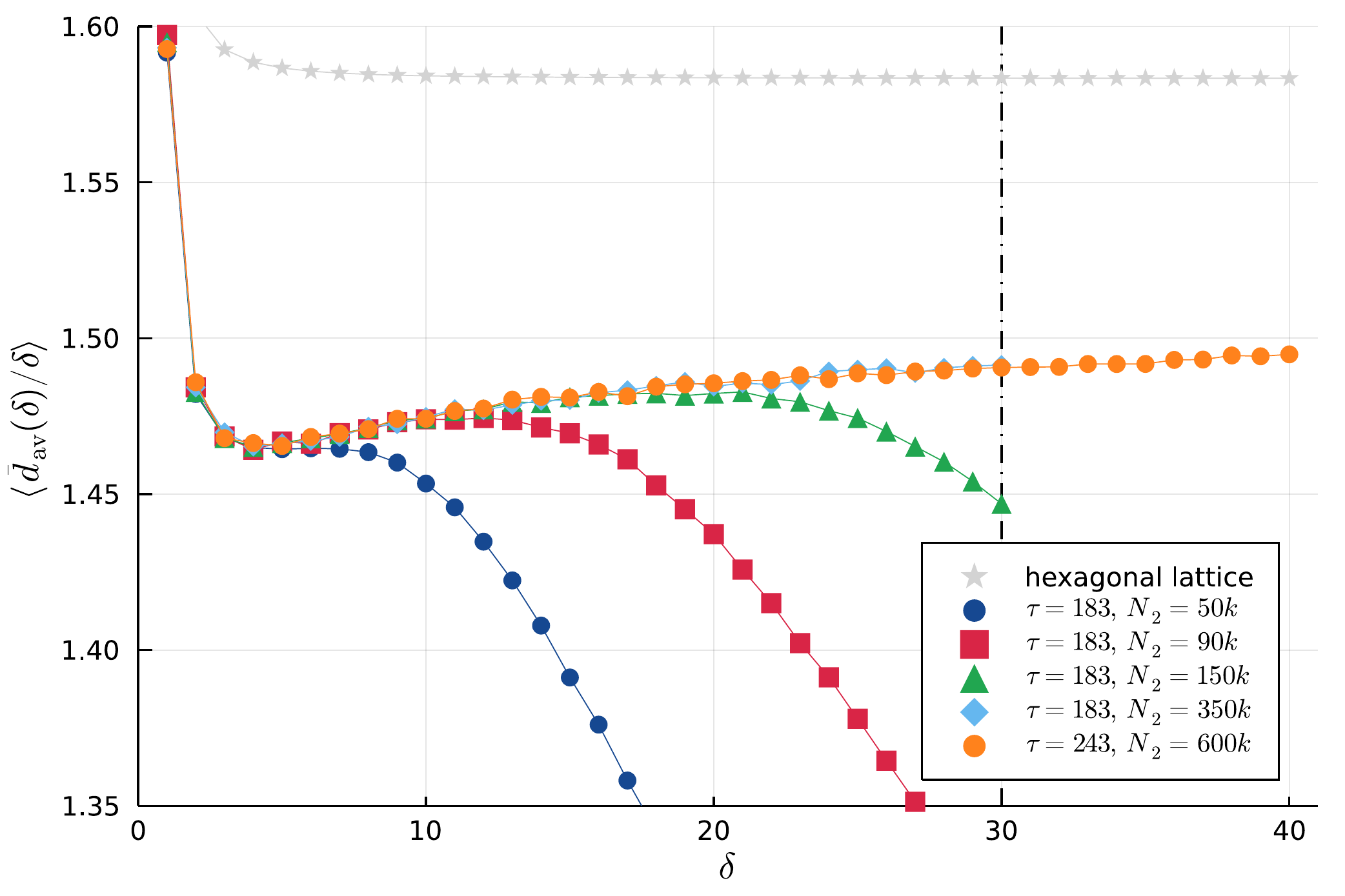}
	\caption{Curvature profile $\langle \bar{d}_{\rm av}(\delta)/\delta \rangle$ as a function of the link distance $\delta$ for
	CDT quantum gravity on a two-torus, for various volumes $N_2$ and time extensions $\tau$.  Any data for $\tau\! =\! 183$ lying on or to the left of the dash-dotted vertical line at $\delta\! =\! 30$ is not influenced by the periodic identification of time. The classical curvature profile 
	of the flat hexagonal lattice is shown for comparison. (Error bars are smaller than dot size.)}
	\label{fig:ric2dcdt}
\end{figure}

Our results for the curvature profile $\langle \bar{d}_{\rm av}(\delta)/\delta \rangle$ as a function of the link distance 
are shown in Fig.\ \ref{fig:ric2dcdt}. The five data sets for the CDT ensemble display an almost identical behaviour for $\delta\!\leq\! 5$, namely,
a steeply falling curve due to short-distance lattice artefacts, familiar from previous investigations \cite{qrc1,qrc2,qrc3,qrc4}, both in a classical
and a quantum context. Beyond this region, all interpolated curves show a gentle incline, which for the three smaller volumes continues up to some maximum, 
beyond which they decrease rather fast. The location of the maximum appears to be roughly proportional 
to the volume $N_2$, that is, effectively to the volume of the spatial slices. 
We have not included any data points that are potentially affected by the choice of the compactified boundary condition
in the time direction, which for $\tau\! =\! 183$ would lie to the right of $\delta\! =\! 30$, marked by the vertical line in Fig.\ \ref{fig:ric2dcdt}. The simulation data
for the two larger volumes do not exhibit any downward slope in the investigated $\delta$-ranges.  

\begin{figure}[t]
	\centering
	\includegraphics[width=0.6\textwidth]{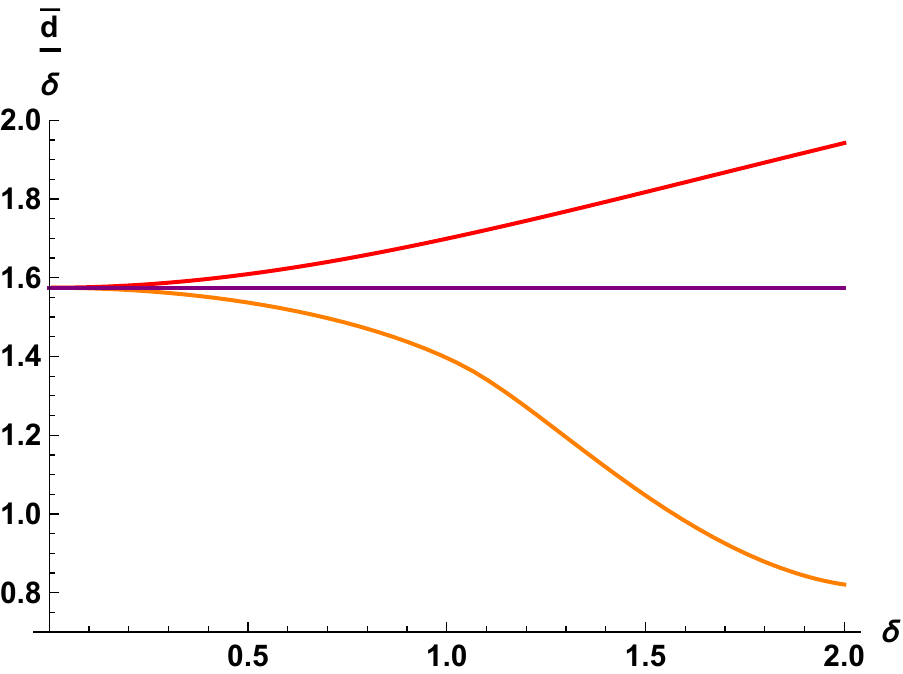}
	\caption{Classical curvature profiles $\bar{d}_{\rm av}(\delta)/\delta$ as a function of the distance $\delta$ for
	two-dimensional Riemannian manifolds of constant curvature: a hyperboloid of curvature radius 1 (top), flat space (middle) and a sphere of
	curvature radius 1 (bottom).  }
	\label{fig:riccishort}
\end{figure}

In trying to interpret these outcomes, let us recall the behaviour of the curvature profiles on constantly curved Riemannian spaces \cite{qrc1}, 
which for small $\delta$ is also captured by the expansion (\ref{2dexp}). 
Considering that the quantum geometry we are investigating is presumably homogeneous (``looks the same everywhere") on sufficiently large
scales, classical spaces of constant Gaussian curvature are natural spaces to compare with. 
As illustrated by Fig.\ \ref{fig:riccishort}, 
their curvature profiles increase for negative and decrease for positive curvature as a function of $\delta$, and are constant
(in fact, for all $\delta$) on flat space. However, what we observe in the measurements 
is not an obvious match for any of these cases. The fall-off behaviour found for the smaller
volumes could be an indication of positive curvature, of a topological effect due to the compact spatial boundary condition, or of a combination of both. 
Alternatively, if we can demonstrate that
the fall-offs are a purely topological effect, discarding the corresponding data points for large $\delta$ could plausibly lead to a single, universal
enveloping curve, roughly along the data points taken at the largest volume $N_2\! =\! 600k$ in Fig.\ \ref{fig:ric2dcdt}, whose
upward slope may indicate the presence of (a small) negative curvature. 

On general grounds, if the impact of the global torus topology can be quantified and removed from our data, the remainder should describe
the pure, quasi-local geometry of a (possibly infinitely extended) contractible region of spacetime, without any topological or boundary effects. 
However, note that such a scenario lacks a distinguished macroscopic scale, which makes it difficult to understand how its curvature behaviour 
could possibly resemble that of
a constantly curved space of either positive or negative curvature, both of which are characterised by a curvature radius $\rho$ 
(which in Fig.\ \ref{fig:riccishort} is set to 1). 
In terms of a classical interpretation, this only seems to leave flat space. For example, one might envisage a situation
where positive and negative curvatures on microscopic scales average out to produce an effectively
flat space on sufficiently large scales. 
On the other hand, since we are dealing with a pure quantum system, as already emphasized in the introduction, 
there are no compelling arguments for why the quantum geometry should be flat.\footnote{Note that the quantum Ricci curvature for
$\delta\! >\! 1$ is distinct from the deficit-angle prescription of Regge calculus \cite{regge}, summed over a ball of radius $\delta$,
and in particular does not obey a Gauss-Bonnet theorem in two dimensions. 
Regge's definition of local curvature on $D$-dimensional triangulations in terms of the number of $D$-simplices meeting at a
$(D-2)$-simplex is a valid notion for finite simplicial manifolds, but does not give rise to a well-defined renormalized curvature  
in the nonperturbative quantum theory, unlike the quantum Ricci curvature \cite{qrc1,qrc4}.} 

Let us nevertheless examine the classical flatness hypothesis a bit further. Is it possible that the enveloping curve of Fig.\ \ref{fig:ric2dcdt}
would flatten out if we added data for larger volumes? In other words, could there be lattice artefacts 
that mask a ``true" constant behaviour of the data points in the range measured up to now? 
To assess this issue, we have computed the curvature profile 
$\bar{d}_{\rm av}(\delta)/\delta$ of the regular triangulation of the (infinitely extended) flat plane, corresponding to a hexagonal lattice, as a function of
the link distance, up to $\delta\! =\! 40$.
Using translation invariance, the double sum over points $p$ and $p'$ in eq.\ (\ref{avdisttri}) reduces to a single sum over points $p'$ at distance $\delta$ 
from a fixed point $p$. The geometry of the situation is illustrated by Fig.\ \ref{fig:cdt-sphere-pair} for $\delta\! =\! 2$, where one should still average over all locations
$p'$ that lie on the blue ``sphere" $S_p^\delta$. We have included the exactly computed values of the classical curvature profile for the hexagonal lattice for 
$\delta\!\geq\! 3$ in Fig.\ \ref{fig:ric2dcdt}, for comparison with the simulation data. After an initial ``overshoot" caused by lattice artefacts, 
where the points follow a monotonically decreasing curve, they settle to an approximately constant value for
$\delta\! \gtrsim\! 6$, without going through a dip, as the quantum measurements do. (Recall that the offset in the $y$-direction of the curvature profile
is a lattice-dependent, nonuniversal quantity.) This agrees with the expectation that the quantum Ricci curvature vanishes identically for sufficiently
large $\delta$ and is in line with earlier investigations of flat lattices and of approximations of flat space by Delaunay triangulations \cite{qrc1}.\footnote{Our data 
for the hexagonal lattice differ slightly from those presented in \cite{qrc1}, which did not include an average over all directions.}
We conclude that comparison with a classical flat lattice does not provide support to the interpretation that the quantum geometry of the two-torus
is effectively flat. An additional argument against the existence of a $\delta$-value far away from the cutoff scale, above which the curve for
$\langle \bar{d}_{\rm av}(\delta)/\delta \rangle$ would flatten out, is the absence of a distinguished macroscopic physical scale, as already pointed out
earlier.   

\begin{figure}[t]
	\centering
	\includegraphics[width=0.7\textwidth]{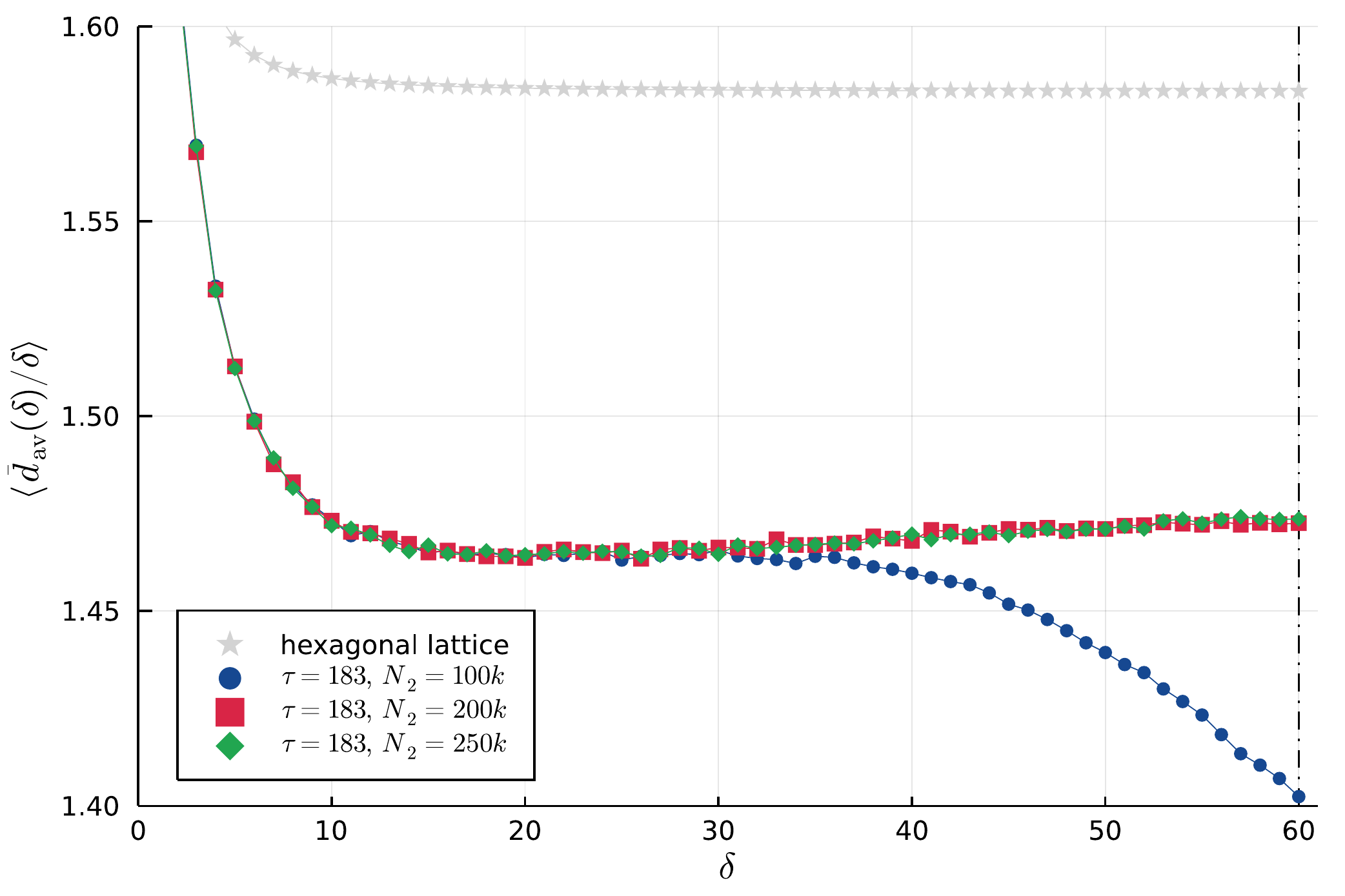}
	\caption{Curvature profile $\langle \bar{d}_{\rm av}(\delta)/\delta \rangle$ as a function of the dual link distance $\delta$ for
	CDT quantum gravity on a two-torus, for various volumes $N_2$ and time extension $\tau\! =\! 183$. The dash-dotted vertical line at $\delta\! =\! 60$ 
	marks the upper end of the $\delta$-range not influenced by the periodic identification of time. The classical curvature profile 
	of the flat hexagonal lattice as a function of the dual link distance is shown for comparison. (Error bars are on the order of the dot size.)}
	\label{fig:dric2dcdt}
\end{figure}

As a cross-check of our measurements in terms of the link distance, 
we also measured the curvature profile $\langle \bar{d}_{\rm av}(\delta)/\delta \rangle$ using the dual link distance.  
By universality, one would expect to find similar results, up to a finite rescaling of $\delta$ as discussed earlier. This is borne out
by our measurements, which are presented in Fig.\ \ref{fig:dric2dcdt}, together with the corresponding classical curvature profile of the 
hexagonal lattice for comparison. The three depicted data sets coincide initially, including the rapid fall-off for small $\delta$ and subsequent
slight dip, before the curve for the smallest volume starts decreasing, while those for the two larger volumes continue on a gently increasing
slope. The behaviour is qualitatively very similar to that of Fig.\ \ref{fig:ric2dcdt}, up to a rescaling of roughly a factor 2 along the $x$-axis 
(regarding the location of the minimum and the onset of the fall-off of the curves), and
a reduction of the amplitude of the curves in the $y$-direction. 

Before we reach a final conclusion on the interpretation of the measured curvature profiles of Figs.\ \ref{fig:ric2dcdt} and \ref{fig:dric2dcdt}, 
we still need to understand and quantify to what extent
the periodic identification of the spatial direction of the torus geometries influences the quantum Ricci curvature measurements. 
Because of the fluctuating nature of the spatial slices in the quantum ensemble, this analysis is not completely straightforward. 
It will be the subject of the following section.

\begin{figure}[t]
	\centering
	\includegraphics[width=0.7\textwidth]{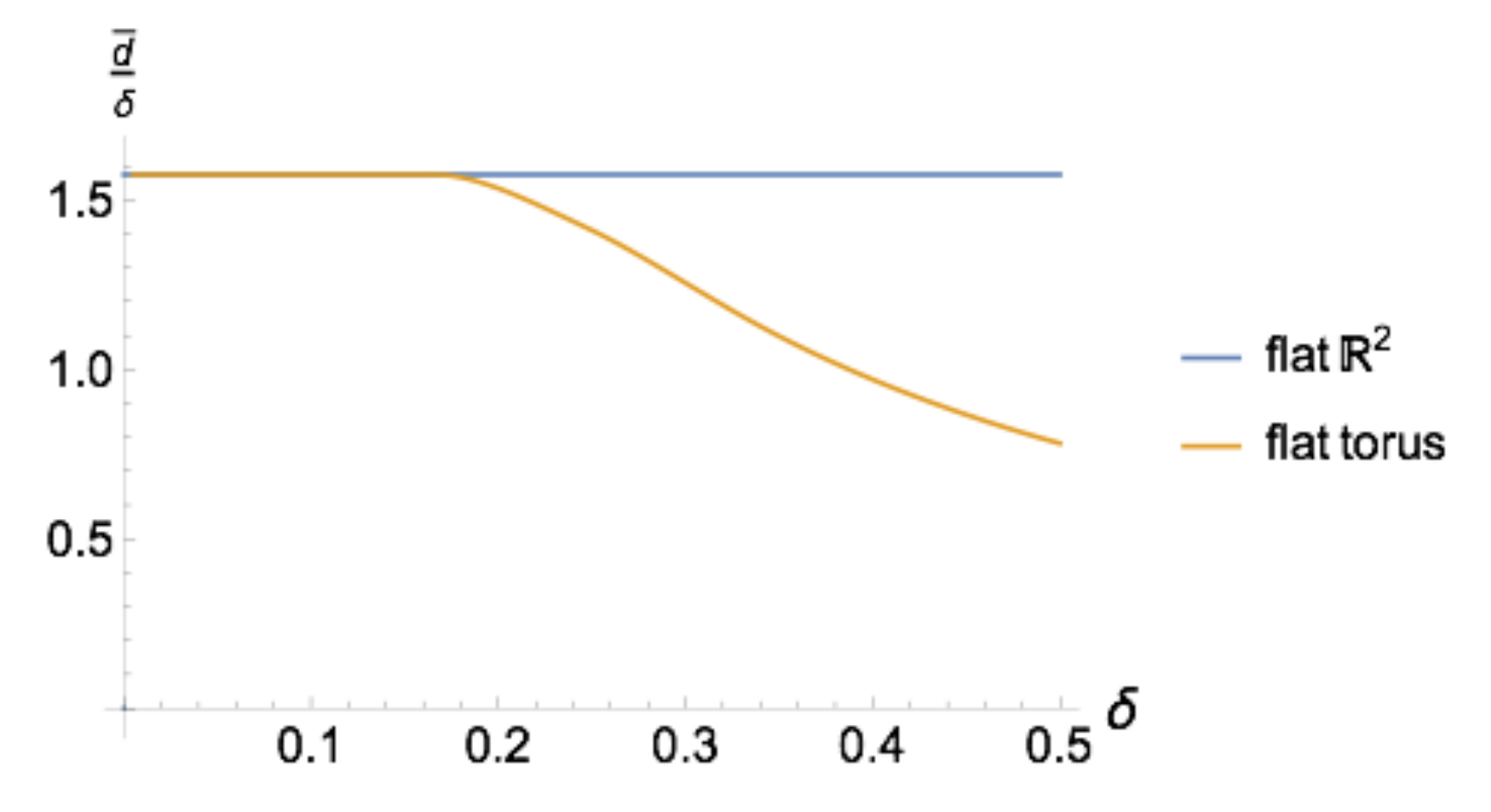}
	\caption{Normalized average sphere distance $\bar{d}/\delta$, measured in one of the directions of a flat classical 
	unit torus.  For comparison, we have included the same quantity on a flat space without compactifications. The two curves coincide initially, but for
	$\delta\! \gtrsim\! 0.17$ the torus data stop following a horizontal curve and switch abruptly to a quasi-linear
	decreasing behaviour. [Figure courtesy of N.\ Klitgaard.]}
	\label{fig:flattorus}
\end{figure}

\section{The influence of global topology} 
\label{sec:topology}

The effect of the global topology on the curvature profile of a smooth classical manifold can be illustrated neatly by considering a flat 
torus \cite{thesisklitgaard}.
The example we consider is a two-torus, obtained by the pairwise identification of the opposite sides of a flat square parametrized by $x\!\in\! [0,1]$
and $y\!\in\! [0,1]$. Fig.\ \ref{fig:flattorus} shows the normalized average sphere distance $\bar{d}(S^\delta_p,S^\delta_{p'})/\delta$ corresponding to 
a point pair $(p,p')$, where
$p\! =\! (0.5,0.5)$ and $p'\! =\! (0.5+\delta,0.5)$ have a geodesic distance $\delta\!\in\! [0,0.5]$ purely along the $x$-direction. 
By virtue of the triangle inequality, the largest possible distance between any pair of points $(q,q')\!\in\! S^\delta_p\!\times\! S^\delta_{p'}$ 
is given by $3\delta$, which for sufficiently small $\delta$ is realized by $(q,q')\!=\! ((0.5\! -\!\delta,0.5),(0.5\! +\! 2\delta,0.5))$.
However, if $\delta\! >\! 1/6\!\approx\! 0.1666$, there is a shorter geodesic between $q$ and $q'$ that runs around the torus in the opposite direction, 
leading to the shorter distance $d_g(q,q')\! =\! 1\! -\! 3\delta$. The fact that there is more than one geodesic between a given pair of
points, despite the local flatness of the manifold, is a consequence of the nontrivial topology of the torus. More specifically, there is a
two-parameter family of geodesics between each pair of points, which cannot be smoothly deformed into each other and which differ by their
relative winding numbers around the two torus directions. Of course, only the shortest of these geodesics determines the distance $d_g(q,q')$
that enters into the computation of the average sphere distance. This should be compared with the situation in the flat plane, where there is always a
unique geodesic between two points, determining their Euclidean distance. 

The two cases are contrasted in Fig.\ \ref{fig:flattorus},
where the normalized average sphere distance on flat $\R^2$ is a constant $\approx\! 1.5746$ (c.f.\ eq.\ (\ref{2dexp})). The corresponding curve for the torus, 
computed with the help of Monte Carlo simulations \cite{thesisklitgaard}, coincides with the horizontal curve for $\delta\!  \lesssim\! 1/6$, and beyond
this point starts deviating sharply, entering a regime of quasi-linear decline. In other words, beyond a finite threshold value for $\delta$, there is
a strong influence of the global topology on the average sphere distance, which is systematically lowered due to the presence of ``shortcuts"
between point pairs $(q,q')$ that do not exist for trivial topology. This purely topological effect can be compared to the effect of nontrivial geometry,
which for a sphere of constant curvature also leads to a decreasing curvature profile. However, as illustrated by the corresponding  
curve in Fig.\ \ref{fig:riccishort}, the downward slope starts as soon as $\delta\! >\! 0$, as described by eq.\ (\ref{2dexp}), and without an initial flat plateau. 

\begin{figure}[t]
	\centering
	\includegraphics[width=0.7\textwidth]{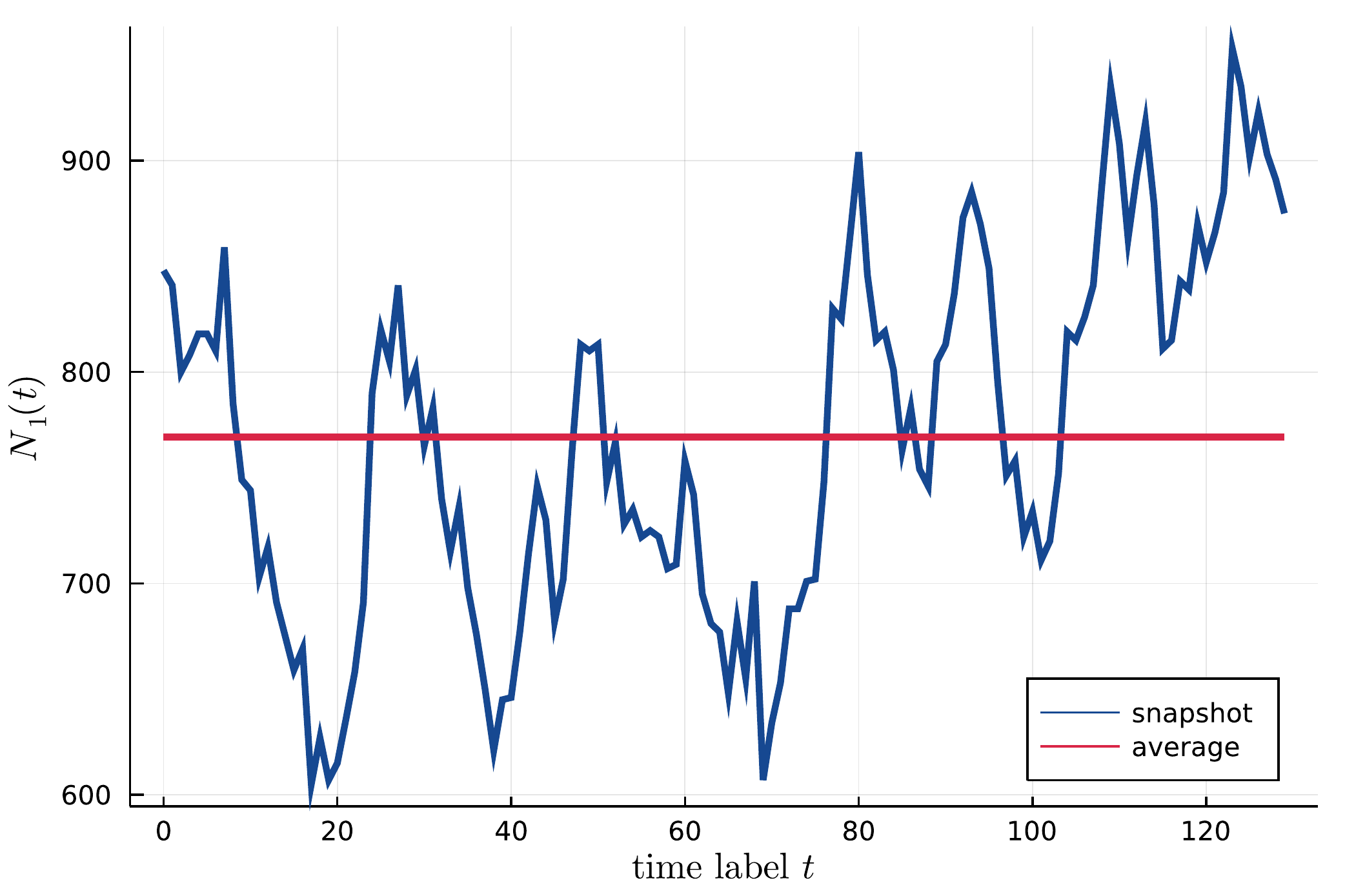}
	\caption{Volume profile $N_1(t)$ of a typical torus geometry in the CDT ensemble, together with the average $\bar{N}_1$ (red), for time
	extension $\tau\! =\! 130$ and volume $N_2\! =\! 200k$.}
	\label{fig:volume-profiles}
\end{figure}

Returning to quantum gravity, the sliced structure and fixed time extension of the CDT geometries make it straightforward to deal 
with the periodic boundary condition in the time direction in the same way as one would on a classical torus. 
To avoid any influence of this (unphysical) choice on the measurement of the curvature profile, it is sufficient to observe the bound
$\delta\! \leq \!\tau/6$. The effect of the compactness in the spatial direction is much more difficult to assess because the discrete volume 
(number of edges) $N_1(t)$ of a spatial slice at discrete proper time $t$ is subject to large quantum fluctuations. This happens because there 
is only a single
length scale in two-dimensional quantum gravity, set by $1/\sqrt{\lambda}$, where $\lambda$ is the cosmological constant or,
for fixed two-volume, by $\sqrt{N_2}$. The expectation value of $N_1(t)$ and its standard deviation are therefore of the same order
of magnitude. A snapshot of the volume function $N_1(t)$, $t=0,1,\dots ,\tau\!-\! 1$ from a Monte Carlo simulation with $\tau\! =\! 130$ and $N_2\! =\! 200k$
is shown in Fig.\ \ref{fig:volume-profiles}, together with the average spatial volume $\bar{N}_1(t) \! =\! N_2/(2\tau)$ (keeping in mind that 
$N_2\! =\! 2\sum_t N_1(t)$). Potentially problematic for the curvature measurements are path integral configurations that contain spatial
slices of very small volume, on the order of $6\delta$ and below. However, 
the presence of fluctuations means that for fixed time $t$ there is no strict lower bound\footnote{at or above the kinematical 
minimum $N_1\! =\! 3$ required for a simplicial manifold} on the spatial volumes $N_1(t)$ that can occur. 
Moreover, even exact knowledge of
the distribution of $N_1$ in the ensemble would not be sufficient to determine how the sphere distance measurements are affected, 
because the double-spheres extend over many time slices, as will typical shortcuts between pairs of points. 

In practical terms, we will address this challenge by numerically establishing an upper bound $\delta_{\rm max}$ on $\delta$ for given 
$N_2$ and $\tau$, such that ``topological shortcuts" occur very rarely, i.e.\ at a prescribed low rate, and have a negligible
influence on the curvature profile. 
This analysis requires a prescription of how to define such shortcuts and find them on a given geometric configuration, an issue we turn to next.    

\subsection{Intersection number of geodesics}
\label{inter}

\begin{figure}[t]
	\centering
	\begin{subfigure}[t]{0.45\textwidth}
	\centering
		\includegraphics[height=0.8\textwidth]{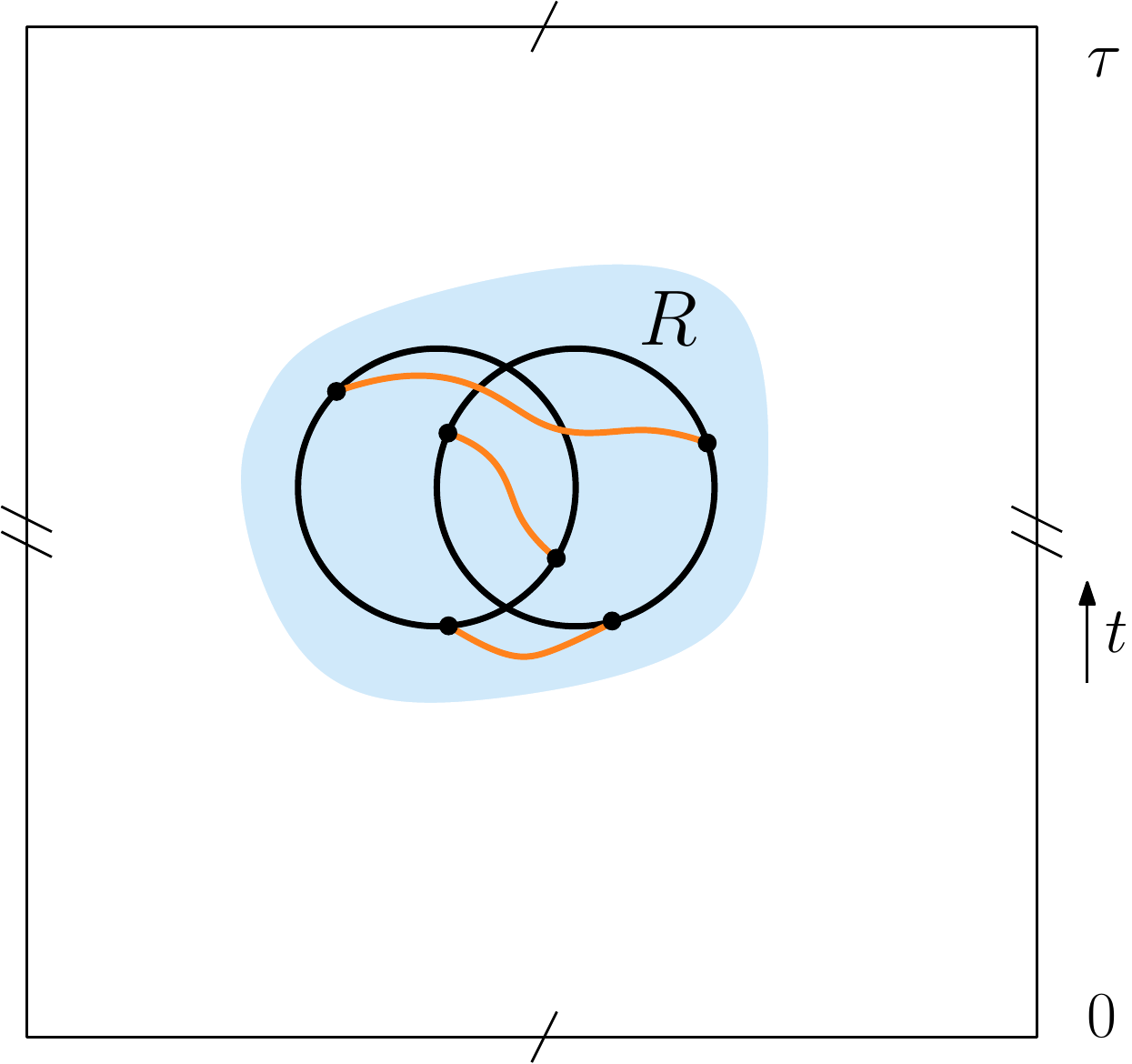}
		\caption{
		}
		\label{fig:torus-topo-safe}
	\end{subfigure}
	\hfill
	\begin{subfigure}[t]{0.45\textwidth}
	\centering
		\includegraphics[height=0.8\textwidth]{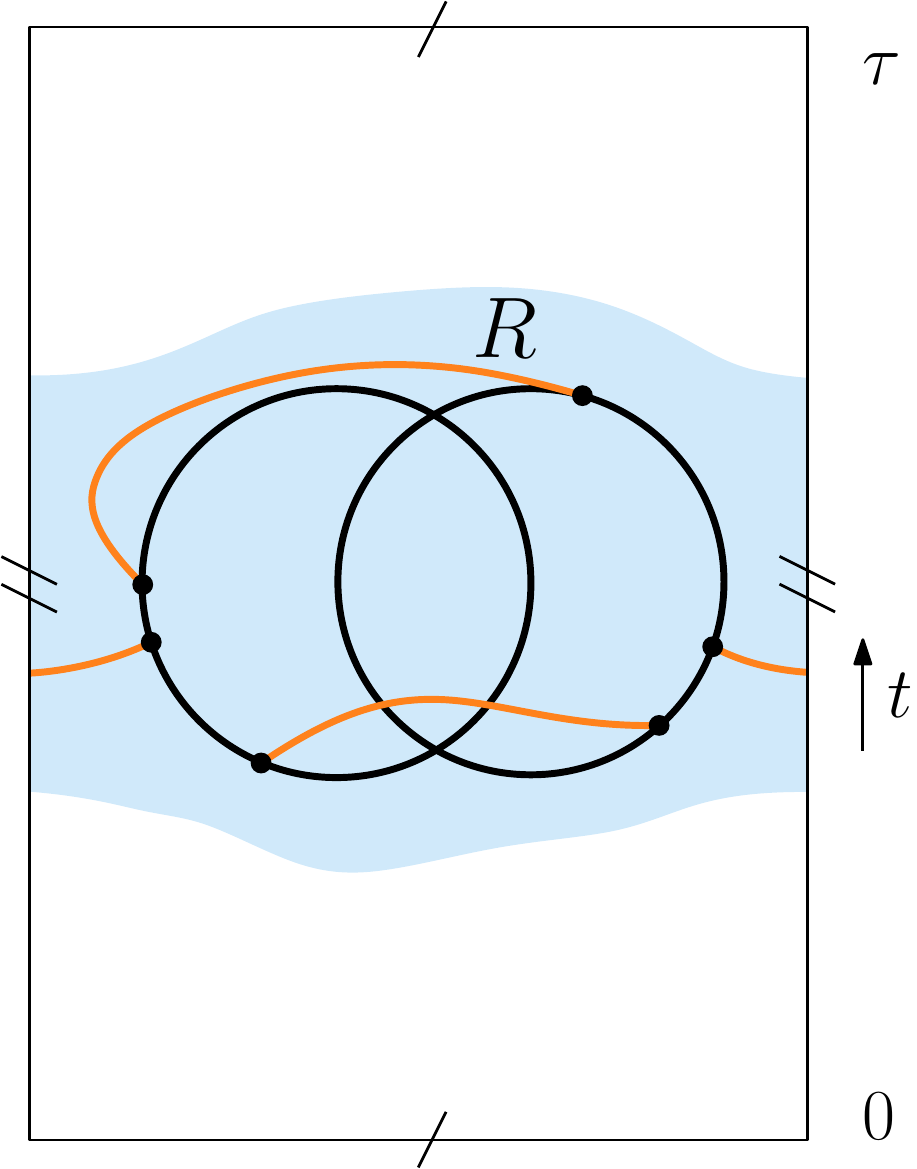}
		\caption{
		}
		\label{fig:torus-topo-unsafe}
	\end{subfigure}
	\caption{Schematic representation of a circle pair on a toroidal CDT geometry, including some shortest geodesics
	between pairs of points from the two circles. In the regular case (no effect of global topology) all shortest geodesics are contained 
	in a disc-shaped region $R$ (left). If the spatial extension becomes too short compared to the diameter of the double circle, not all
	shortest geodesics are contained in a contractible disc (right). This topological effect affects the computation of the average
	sphere distance.
	}
	\label{fig:torus-topo}
\end{figure}

Fig.\ \ref{fig:torus-topo} illustrates the geometric situation we want to analyze, consisting of a pair of spheres (circles) $S_p^\delta$, $S_{p'}^\delta$ 
of radius $\delta$ located somewhere on a curved CDT toroidal geometry of time extension $\tau$ (vertical direction) and with spatial slices   
of fluctuating size $N_1$ (fluctuations not shown), both periodically identified as indicated. We assume $\delta\! \leq\! \tau/6$, 
such that the time periodicity does not affect the
average sphere distance. If the size of all spatial slices is large enough, every geodesic between a point $q\!\in\! S_p^\delta$ and a point
$q'\!\in\! S_{p'}^\delta$ is completely contained in a simply connected region $R$ that includes the two circles and the union of their interiors (blue
region in Fig.\ \ref{fig:torus-topo-safe}). By contrast, in Fig.\ \ref{fig:torus-topo-unsafe} some of the spatial slices are too small, in the sense
that at least for some point pairs $(q,q')$ the shortest geodesic crosses the vertical line along which the torus has been cut open. In this case, a
region $R$ containing both the circles and all shortest geodesics between pairs $(q,q')$ cannot be disc-shaped, but instead must wind around the
spatial direction of the torus, forming a noncontractible annulus.   

Given a CDT geometry $T$ and a pair of circles $(S_p^\delta, S_{p'}^\delta)$
on it, the key idea for monitoring the occurrence of this latter phenomenon is to consider a suitable
closed curve $\gamma$ in $T$ that winds around the time direction once and lies {\it outside} the union $C$ of the circles and their interiors, defined by 
$C\! :=\! \left\{q \,|\, d(q, p) \leq \delta \lor d(q,p') \leq \delta\right\}$. Then, if no shortest geodesic between any point pair 
$(q,q')\in S_p^\delta \times S_{p'}^\delta$ ever crosses $\gamma$, we are dealing with the ``regular" case depicted in Fig.\ \ref{fig:torus-topo-safe},
which is unaffected by the periodic identification of the spatial slices. In fact, this criterion is slightly too restrictive; if $\gamma$ passes near
$C$, it can happen that a shortest geodesic crosses it twice (or an even number of times) in opposite directions, 
despite being part of a perfectly regular configuration. In this case, a continuous deformation of the curve $\gamma$ would eliminate the intersections.
This suggests that we should not simply count the number of intersections, but work with oriented curves  
and their corresponding (oriented) intersection number. 

In the Riemannian case, the intersection number $c_p(\alpha,\beta)$ of two parametrized curves $\alpha(s)$ and $\beta(t)$ at an isolated intersection point $p$
is 1 ($-1$) if the ordered pair of tangent vectors $(\dot{\alpha},\dot{\beta})$ forms a right-handed (left-handed) basis of the tangent space at $p$.
On a triangulation $T$, where geodesics between pairs of vertices by definition follow the links of $T$, it is convenient to choose the reference curve
$\gamma$ along dual links, to avoid any nontrivial overlaps of the curves along links (see Fig.\ \ref{fig:crossing-primal-dual} below). 
More details on the choice and construction of the curve $\gamma$ can be found in the Appendix. 

\begin{figure}[t]
	\centering
	\begin{subfigure}[t]{0.45\textwidth}
	\centering
		\includegraphics[height=0.8\textwidth]{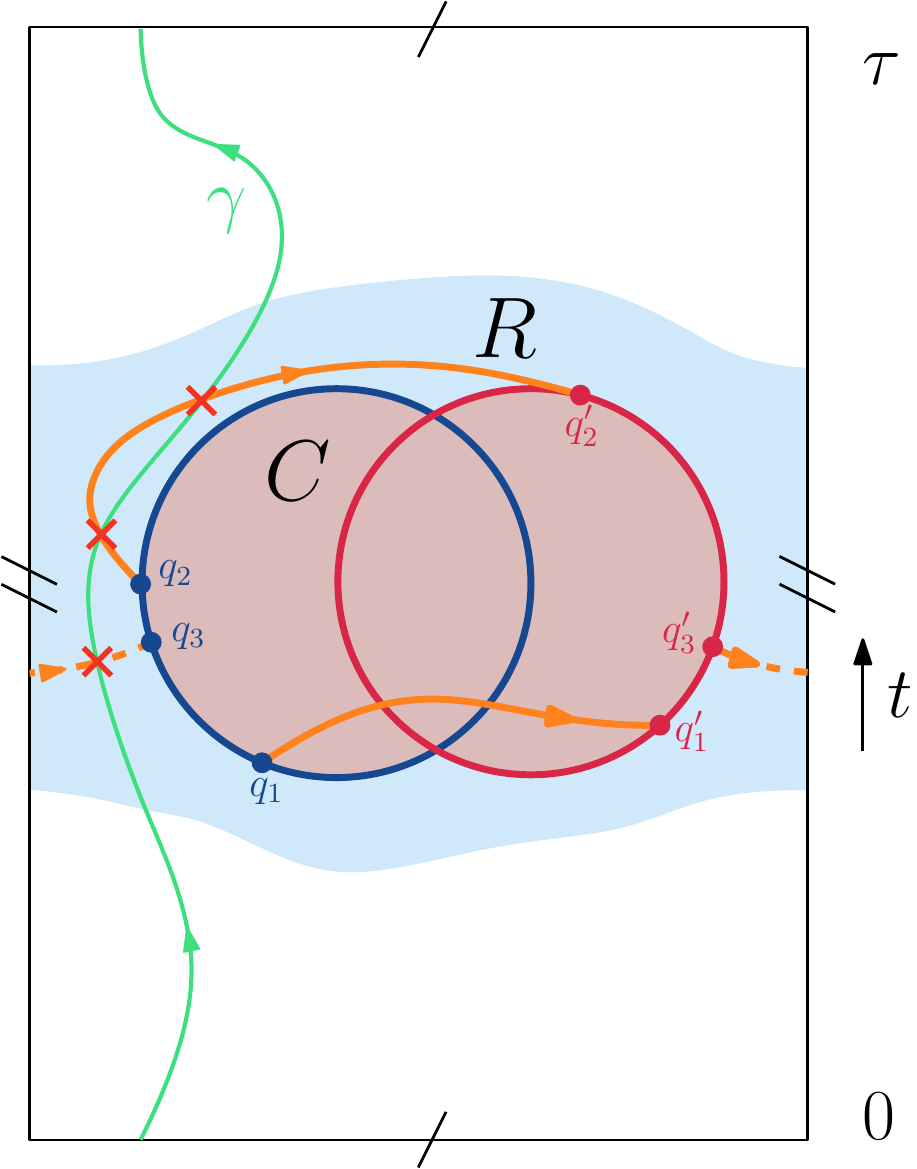}
		\caption{
		}
		\label{fig:torus-topo-cut-unsafe}
	\end{subfigure}
	\hfill
	\begin{subfigure}[t]{0.45\textwidth}
	\centering
		\includegraphics[height=0.8\textwidth]{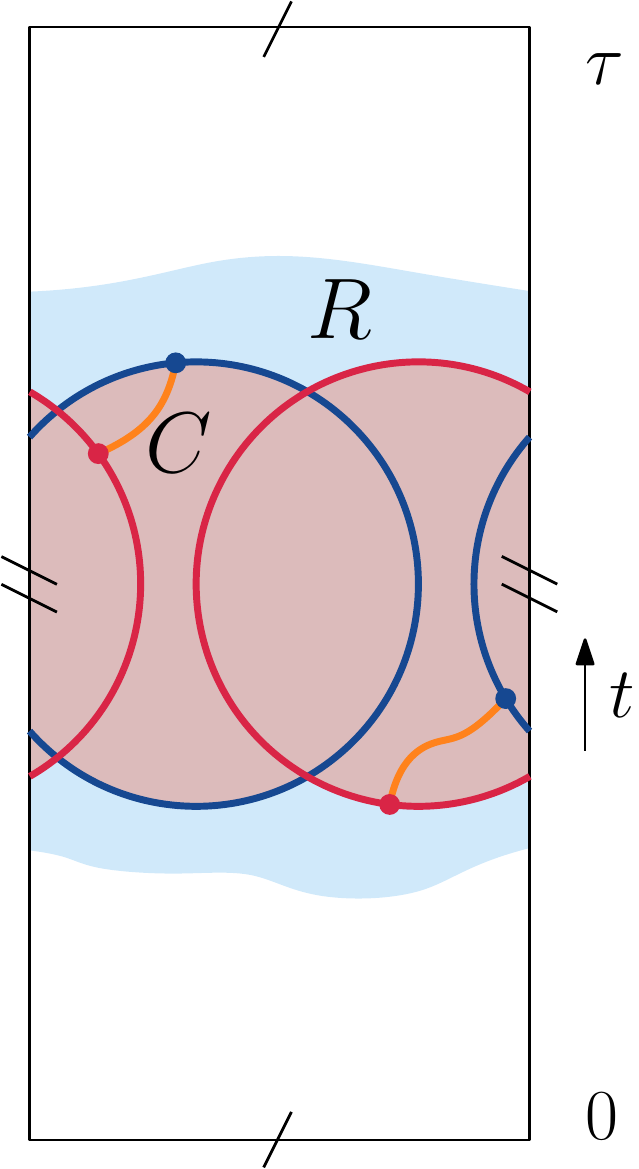}
		\caption{
		}
		\label{fig:torus-topo-wrap}
	\end{subfigure}
	\caption{Configuration of two spheres $S_p^\delta$ (blue) and $S_{p'}^\delta$ (red) on a CDT geometry $T$. Left: schematic illustration of a level-1 violation. 
	We can find a timelike curve $\gamma$ (green) outside the region $C$ (pink), but there exist shortest geodesics with nonzero intersection number 
	(like the dashed orange path), amounting to topological shortcuts. 
	Right: a level-2 violation, where the region $C$
	self-overlaps and no curve $\gamma$ with the required properties exists. The two shortest geodesics shown (orange) represent topological shortcuts. 
	}
	\label{fig:torus-topo-cut}
\end{figure}

Given a shortest geodesic $\phi (q,q')$ between two vertices $q$ and $q'$ on a CDT configuration $T$, oriented to run from $q$ to $q'$, say, 
and a (oriented) reference curve
$\gamma$ along dual links, we define their total intersection number $c(\phi,\gamma)$ as the sum over all intersection points $p$ of the intersection
number at $p$, $c(\phi,\gamma)\! =\! \sum_p c_p(\phi,\gamma)$. Following our reasoning above, if for all shortest geodesics $\phi$ between point pairs
$(q,q')$ from a pair of circles we have $c(\phi,\gamma)\! =\! 0$, we are in the regular situation where the computation of the average sphere distance 
is not affected by the global topology. By contrast, the occurrence of nonvanishing intersection numbers $c(\phi,\gamma)$ for a given pair of
spheres $(S_p^\delta, S_{p'}^\delta)$ indicates the presence of shortcuts. We will call this a ``level-1 violation" of the regular case. An example is 
sketched in Fig.\ \ref{fig:torus-topo-cut-unsafe}, where the shortest geodesic between $q_3'$ to $q_3$ has a nonzero intersection number with the curve $\gamma$.

We still need to discuss a situation that occurs for even larger values of $\delta$, 
where no curve $\gamma$ can be found because the region $C$ wraps around the spatial
direction of the torus and creates a self-overlap, as illustrated by Fig.\ \ref{fig:torus-topo-wrap}. It is clear that in this case there will be many shortcuts,   
affecting the average sphere distance in a significant way. We will call this a ``level-2 violation" of the regular case. 

\subsection{Intersection number measurements in CDT}
\label{measure}
Recall that our objective is to quantify the influence of the global topology, more specifically, of the spatial compactness of the CDT geometries,
on the expectation value of the curvature profile. We have seen that the effect is directly related to the presence of topological shortcuts, which
for increasing $\delta$ lead to ever smaller average sphere distances and therefore a smaller value of the curvature profile, compared with the
case of trivial topology. From the comparison with the classical case (Fig.\ \ref{fig:flattorus}), we expect the effect of the compactification to be large
overall, although the shape of the fall-off in $\langle \bar{d}_{\rm av}(\delta)/\delta \rangle$ will presumably be modulated by the presence of the volume 
fluctuations of the spatial slices. Since we now have a computational tool at hand to check for the presence of shortcuts 
(see appendix \ref{sec:appendix} for details on
how this is implemented on CDT geometries), we can address the following questions. Firstly, can the fall-offs for large $\delta$ we observe in the curvature 
profiles of Figs.\ \ref{fig:ric2dcdt} and \ref{fig:dric2dcdt} be attributed entirely to the compact topology? 
Secondly, from the parts of the curvature profiles where the influence of the
shortcuts is negligible, can we extrapolate to the curvature profile of an infinitely extended quantum geometry, representing a ``quantum plane"?    
As we will argue next, from the data we have collected, which of course are subject to numerical limitations in terms of system size and algorithm efficiency,
our qualified answer to both of these questions is in the affirmative.

Our numerical set-up is the same as for the pure average sphere distance measurements described in Sec.\ \ref{sec:numerical}. Namely, we generate
sequences $\{T_n\}$ of triangulations of fixed time extension $\tau$ and spacetime volume $N_2$, and on each geometry $T_n$ compute for each 
$\delta\!\in\! [1,\delta_{\rm max}]$, with $\delmax \leq \tau/6$, the exact average sphere distance of a pair of spheres of radius $\delta$, located randomly on $T_n$. 
Here, we will perform measurements on 5$k$ different triangulations at given $N_2$ and $\tau$, now enhanced by a ``shortcut finder", which we use as follows.
If for a given pair $(S_p^\delta, S_{p'}^\delta )$ of circles 
at least one point pair $(q,q') \in S_p^\delta \times S_{p'}^\delta$ is found to be connected by a shortest path of nonzero intersection number, 
the corresponding data point for $\bar{d}(S_p^{\delta},S_{p'}^{\delta})/\delta$ is marked as having a level-1 violation. 
If the region $C$ of the circles and their interiors is found to wrap around the spatial direction, with self-touchings or self-overlaps,
the corresponding data point for $\bar{d}(S_p^{\delta},S_{p'}^{\delta})/\delta$ is marked as having a level-2 violation.
After collecting all data for given $\tau$ and $N_2$, we determine the fractions $f_1$ and $f_2$ of data points that suffer from a level-1 or level-2 violation
respectively, as a function of $\delta$. (Note that a level-2 violation of a given pair of circles always implies the presence of a level-1 violation, meaning
that $f_2(\delta)\! \leq f_1\! (\delta),\,\forall\delta$.)

These fractions give us estimates of the frequency with which average sphere distance measurements are affected by topological shortcuts, but
they do not provide direct quantitative information about how much they lower the expectation value $\langle \bar{d}_{\rm av}(\delta)/\delta \rangle$.
However, taking into account the behaviour of the classical torus illustrated in Fig.\ \ref{fig:flattorus}, one would expect that shortly after the onset of level-1
violations, their number (and hence their effect on the average sphere distance) grows rapidly as a function of $\delta$. Since there is no sharp onset
for the presence of shortcuts in the quantum theory, as we have already discussed, we should redefine ``onset" to mean ``larger than some small 
threshold fraction $\tilde{f}_1$". In addition, we can use the level-2 violations as estimates for when the measurements will be strongly affected by shortcuts.
Classically, it is clear that when the spheres start to self-overlap, a large part of the distance measurements $d(q,q')$ will be shortcuts (a rough estimate
would be about one third). Their large impact is confirmed by Fig.\ \ref{fig:flattorus}, where this point corresponds to $\delta\! =\! 1/3$,
about halfway down the slope shown for the torus. 
In an attempt to translate this to a criterion for the quantum theory, one may expect that beyond a small, but nonvanishing
threshold value for the fraction $f_2$, one finds a large influence of topological shortcuts on the curvature profile.

\begin{figure}[t]
	\centering
	\begin{subfigure}[t]{0.48\textwidth}
	\centering
		\includegraphics[width=0.95\textwidth]{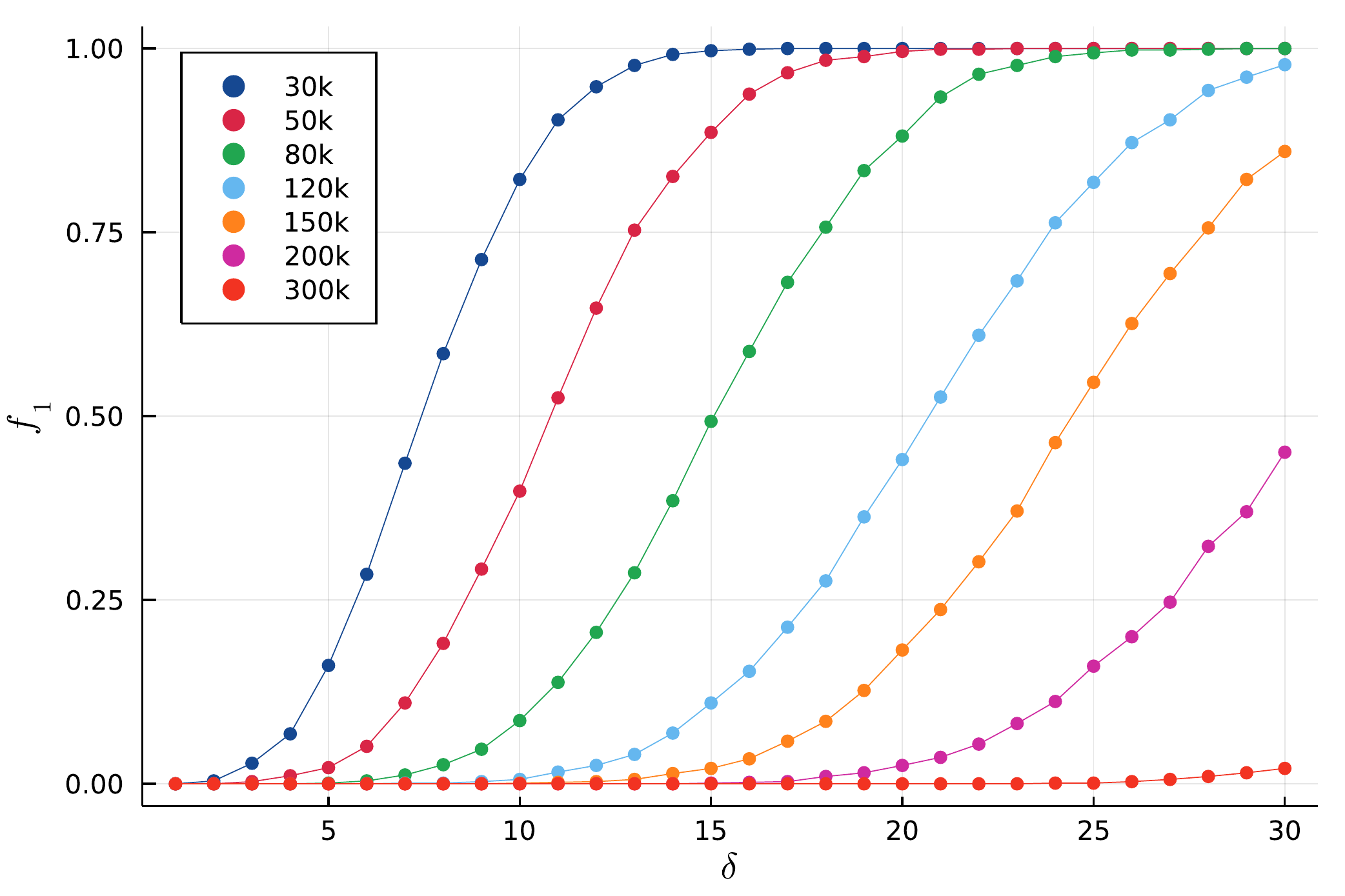}
		\label{fig:violations-primal-lvl-1}
	\end{subfigure}
	\hfill
	\begin{subfigure}[t]{0.48\textwidth}
	\centering
		\includegraphics[width=0.95\textwidth]{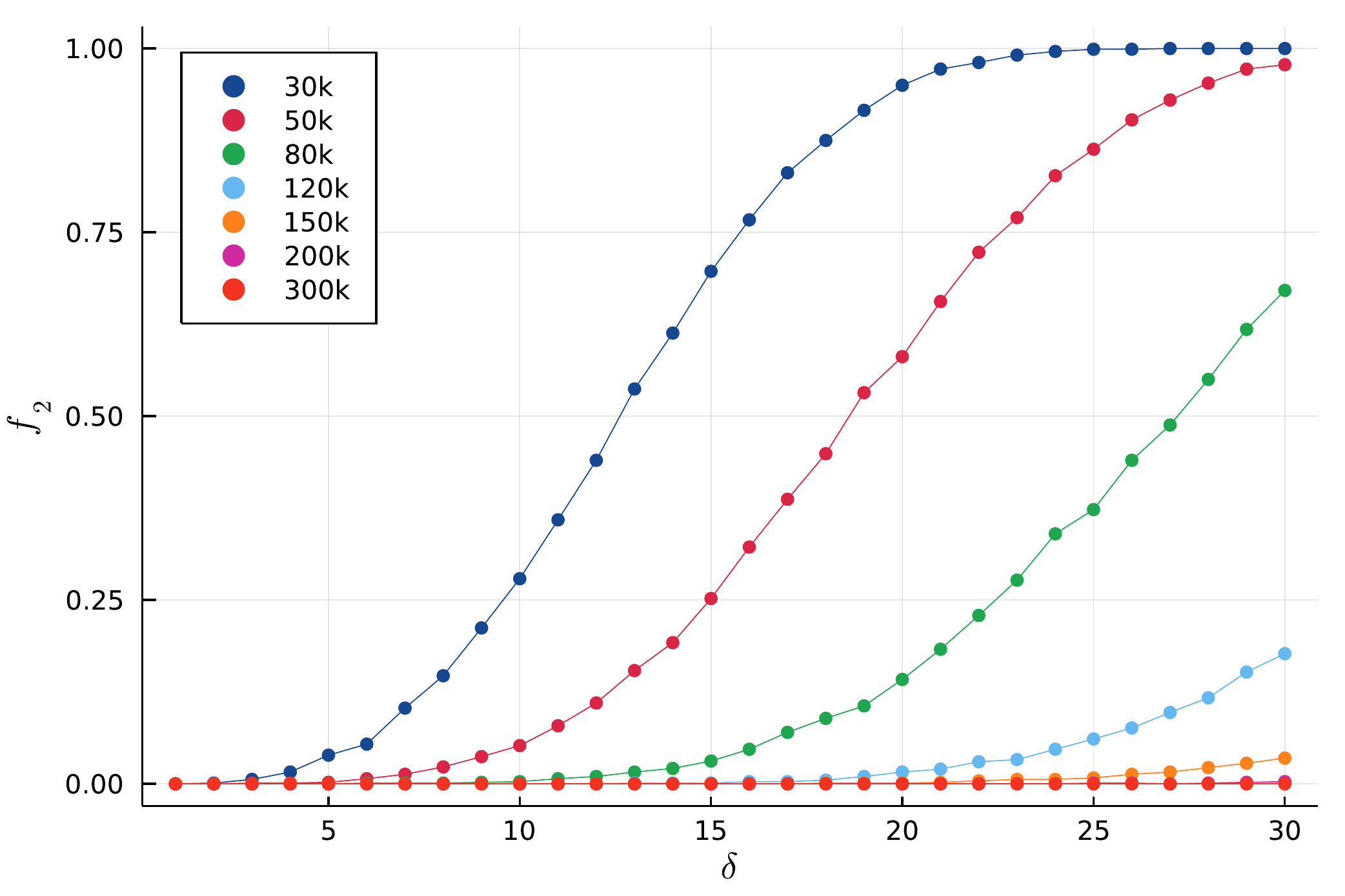}
		\label{fig:violations-primal-lvl-2}
	\end{subfigure}
	\caption{Fraction of level-1 violations $f_1(\delta)$ (left) and level-2 violations $f_2(\delta)$ (right) in CDT simulations with $\tau\! =\! 183$ 
	time slices and volumes $N_2\!\in\! [30k,300k]$.
	}
	\label{fig:violations-primal}
\end{figure}

The data we have found are consistent with these expectations. Our measurements of $f_1(\delta)$ and $f_2(\delta)$ at $\tau\! =\! 183$ and for
a range of volumes are shown in Fig.\ \ref{fig:violations-primal}. 
All curves for $f_1(\delta)$ start out close to zero, stay there for some range of $\delta$ that grows with the volume $N_2$, then undergo a rather rapid rise before
saturating to a value near 1. For the two smallest volumes, there is effectively no $\delta$-range that is not affected by shortcuts, especially considering
that short-distance lattice artefacts extend at least to $\delta\!\approx\! 5$, where anyway one should not trust the average sphere distance measurements. 
At the same time, for the largest system size $N_2\! =\! 300k$ the fraction $f_1$ does not exceed $\sim 0.02$, even at the largest measured value 
$\delta\! =\! 30$, which means that its influence on the curvature profile is very small. The curves for $f_2(\delta)$ have a similar shape, but appear
stretched along the $\delta$-axis by factors between 1.5 and 2.0. 

These findings motivate the introduction of a threshold value $\tilde{f}_1\! =\! 0.01$ for the fraction $f_1$ that characterises the onset of level-1 violations
in the quantum theory. In other words, our working assumption is that average sphere distance measurements whose fraction of shortcuts is below this threshold
are not significantly influenced by the topology. This seems a safe choice, also in view of the fact that at the onset of the shortcut phenomenon
(as $\delta$ increases), the difference in length between a shortcut $\phi(q,q')$ and the corresponding curve with vanishing intersection number between 
$q$ and $q'$ will be small. We have not performed a more detailed analysis comparing the two types of distance for given point pairs $(q,q')$, which is
somewhat involved technically and beyond the scope of this work. 
However, note that the threshold cannot be chosen much larger, because one then quickly reaches $\delta$-values
where level-2 violations become significant (c.f.\ Fig.\ \ref{fig:violations-primal}). For given $N_2$ and $\tau$, the threshold $\tilde{f}_1$ translates
into an upper bound on $\delta$, which we will again denote by $\delta_{\rm max}$.\footnote{Whenever the bound $\delta_{\rm max}$ coming from the time
identification is lower than the one coming from the threshold $\tilde{f}_1$, we implement the lower one of the two.} 
We have checked that for a fixed number of spatial slices
($\tau\! =\! 123$, 183 or 243), $\delta_{\rm max}$ to good approximation scales linearly with the system size $N_2$, as one would expect.

\begin{figure}[t]
	\centering
	\includegraphics[width=0.7\textwidth]{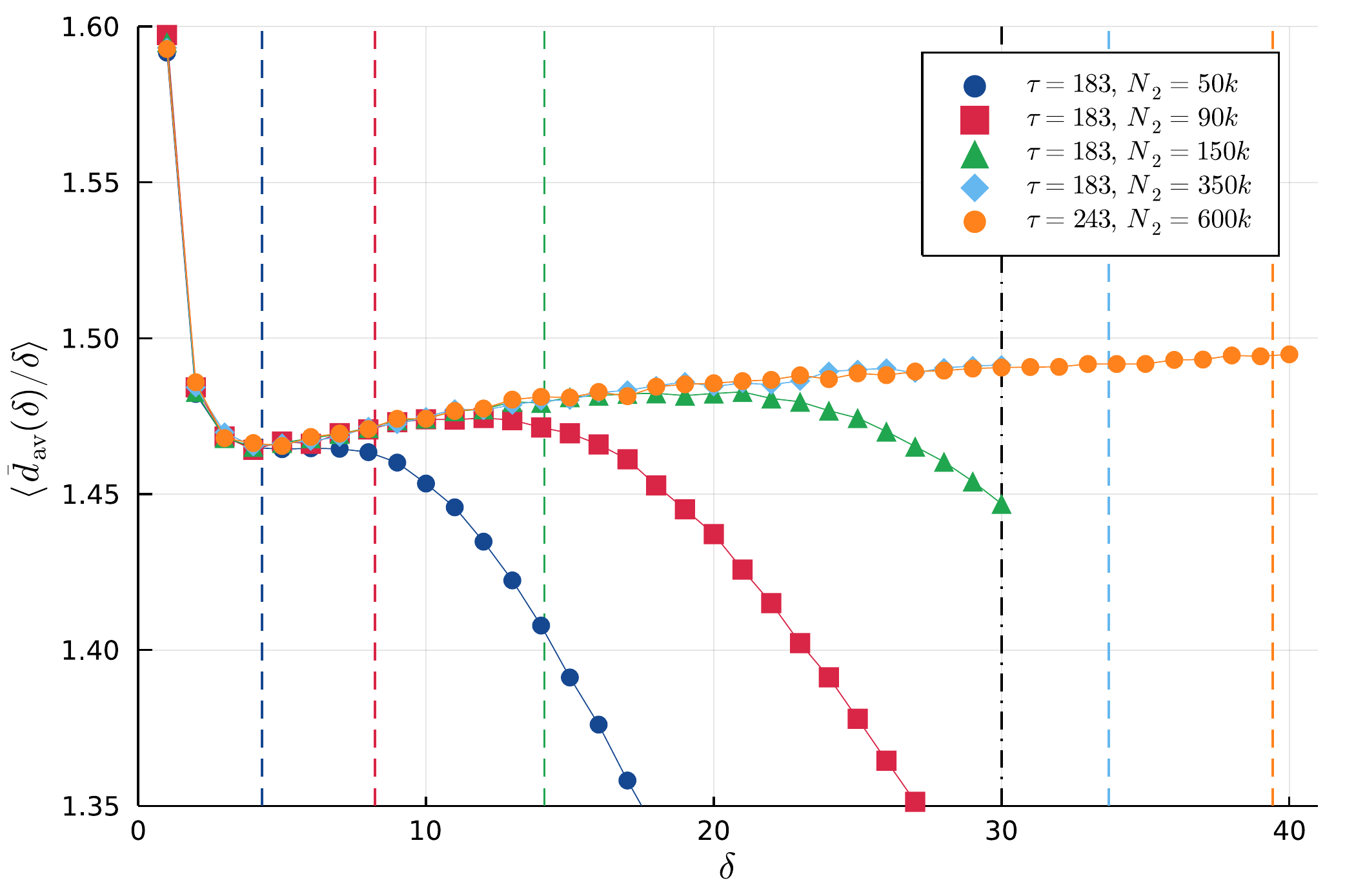}
	\caption{Curvature profiles $\langle \bar{d}_{\rm av}(\delta)/\delta \rangle$ of Fig.\ \ref{fig:ric2dcdt}, supplemented by upper bounds on $\delta$
	(vertical dashed lines), 
	indicating the $\delta$-ranges where topological effects are negligible.}
	\label{fig:ric2dcdt-delmax}
\end{figure}

We will now re-examine the results for the curvature profiles presented Sec.\ \ref{sub:results} in the light of the preceding analysis. We have added to each 
of the curves in Fig.\ \ref{fig:ric2dcdt} the relevant upper bound $\delta_{\rm max}(N_2,\tau)$ in the form of a vertical dashed line, as shown in 
Fig.\ \ref{fig:ric2dcdt-delmax}. 
An immediate observation is that the curvature profile for each of the three smaller volumes only starts sloping downward beyond its associated
bound $\delmax (N_2,\tau)$. This supports the hypothesis that these fall-offs are a direct consequence of the compact spatial topology of the torus, and would not
be present if instead we had studied CDT quantum gravity with spatial slices of the topology of an interval. In addition,
before a curvature profile reaches its bound $\delmax$, it is (within measurement errors) identical to all the other profiles. 
On the basis of the limited range of volumes we have been able to consider, this points to the existence of a common, universal curvature profile in the
limit of infinite volume. 
Interestingly, the resulting, enveloping curve is not a constant, but (beyond the initial dip) increases monotonically and at roughly the same rate 
over the range of $\delta$ we have been able to study. This appears to be a new type of quantum behaviour with no obvious classical analogue,
which we will discuss further in Sec.\ \ref{sec:summary}.

Our current numerical limitations do not allow us to extend the measured profiles to larger $\delta$, to understand whether and at what rate 
the slope continues. On the one hand, 
the computation of the average sphere distance becomes quickly more expensive as $\delta$ increases, slowing down the data collection 
significantly. On the other hand, to accommodate larger values of $\delta_{\rm max}$ the size of the torus would have to be scaled up in both directions,
further increasing simulation times. The combination of these factors makes it difficult to obtain statistics of sufficient quality for link distances much larger than 
the $\delmax\! =\! 40$ we have investigated here.

\subsection{Direction-dependent measurements}
\label{sec:anisotropy}

The quantum Ricci curvature $K_q(p,p')$ defined in eq.\ (\ref{2dexp}) depends on a pair of points $(p,p')$, which can be thought of as a
generalized ``vector", with a certain length (the distance between the points) and ``direction" (along the shortest path between $p$ and $p'$). 
In order to turn it into a proper observable, we integrated $K_q(p,p')$ over all point pairs in the triangulation, subject to the condition 
$d(p,p')\! =\!\delta$, thereby losing any directional information. To nevertheless get an idea of the directional dependence of the quantum Ricci curvature, we 
can investigate separately two different types of ``direction" by referring to the anisotropic, discrete structure of the underlying 
CDT geometries, following the treatment in four dimensions \cite{qrc3}. There is no claim that this construction leads to genuine observables, 
amongst other things, because these geometries do not possess a distinguished time direction. The results should therefore be interpreted with the 
appropriate caution.

\begin{figure}[t]
	\centering
	\includegraphics[width=0.7\textwidth]{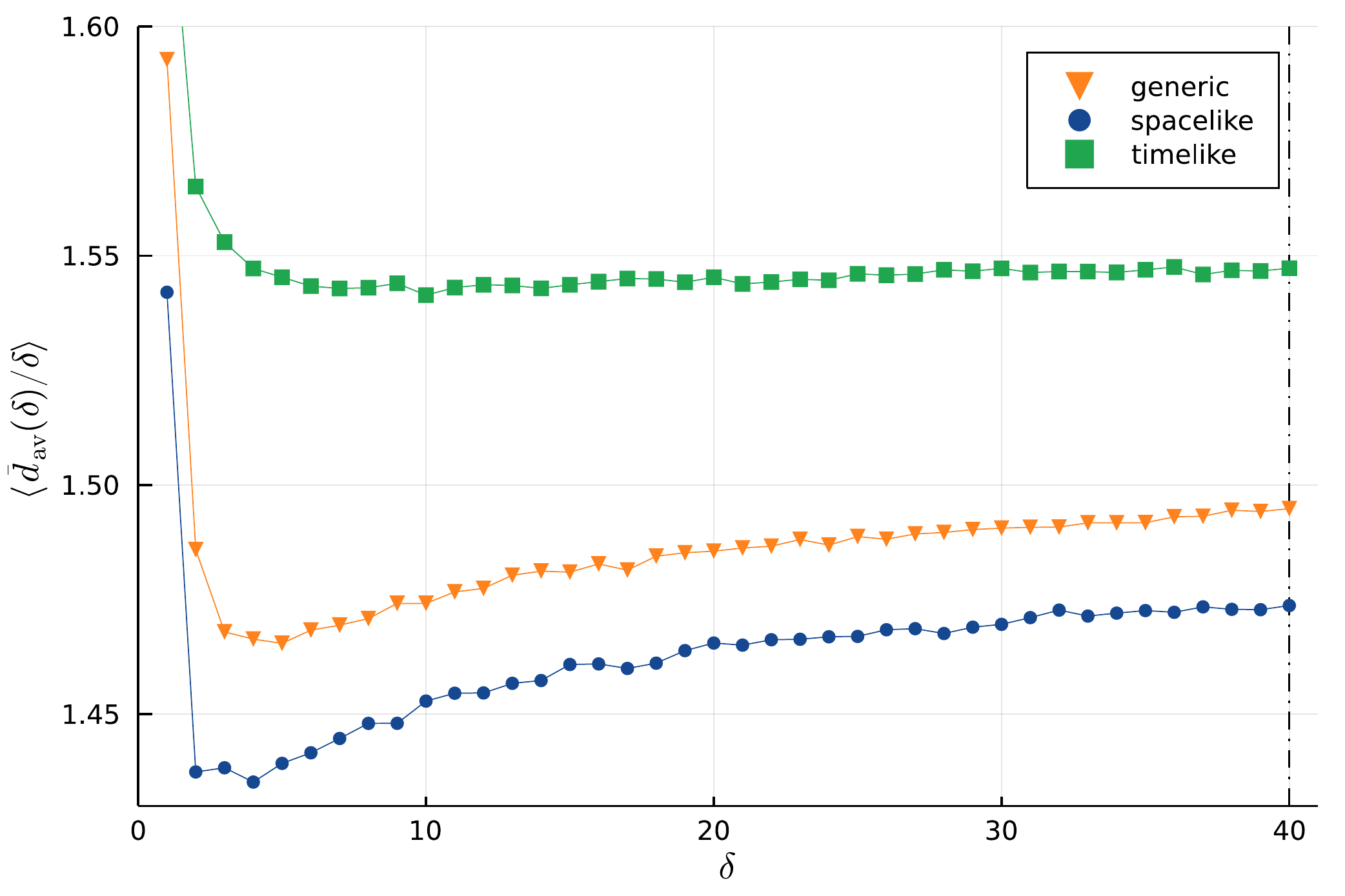}
	\caption{Direction-dependent curvature profiles $\langle \bar{d}_{\rm av}(\delta)/\delta \rangle$ as a function of the link distance $\delta$, for timelike
	(green squares) and spacelike (blue dots) sphere separation, as described in the text, both for $N_2\! =\! 500k$. For comparison, we have added 
	the curvature profile for a generic orientation, measured at $N_2\! =\! 600k$, from Fig.\ \ref{fig:ric2dcdt} (yellow triangles). 	
	(Error bars are on the order of the dot size.)}
	\label{fig:ric-anisotropic}
\end{figure}

We will consider the case where the circle centres $(p,p')$ are separated by $\delta$ spacelike edges contained in the same spatial slice of constant $t$,
and the case where they are separated by $\delta$ timelike edges, such that $\left|t_p\! -\! t_{p'}\right| \! = \! \delta$. The detailed geometry of
the corresponding double-circle configurations differs slightly, due to the anisotropy of the underlying lattice. For example, a timelike separated
circle pair always contains pairs of points with $d(q,q')\! =\! 3\delta$. This is not the case when a circle pair is spacelike separated,
because points $q$ and $q'$ in the same spatial slice can in general be linked by a geodesic which is not contained completely in the same slice. 
As noted in \cite{qrc3,thesisklitgaard}, discretization anisotropies are reflected in different values of the (non-universal)
constant $c_q$ of eq.\ (\ref{2dexp}), corresponding to different vertical offsets in the curve of the normalized average sphere distance. The
interesting question is then whether up to such a relative shift the quantum Ricci curvatures in the space- and timelike directions behave in the same
way, as was observed in CDT quantum gravity in four dimensions \cite{qrc3}.

We have performed $25k$ curvature profile measurements each for timelike and spacelike separations of the circle centres (Fig.\ \ref{fig:ric-anisotropic}).
Both were taken in the range $\delta\! \in\! [1, \delmax\! =\! 40]$ and at volume $N_2\! =\! 500k$, with $\tau\! =\! 243$ for the timelike and $\tau\! =\! 200$
for the spacelike case. For comparison, we have included the direction-averaged data from Fig.\ \ref{fig:ric2dcdt} for $N_2\! =\! 600k$ and $\tau\! =\! 243$.
We observe that the data for the two direction-dependent cases are not completely dissimilar, but are not parallel to each other either. The spacelike
curve resembles that of the direction-averaged data, with an initial dip, followed by a monotonic increase, whose slope however seems to decrease 
as a function of the link distance $\delta$.  
By contrast, the initial dip of the timelike curve is not very pronounced. It is followed by an almost flat regime, with a very slight upward slope. 
The quality of the data is not sufficient to establish whether or not this is compatible with a constant behaviour. 
At this stage, this leaves open the possibility that two-dimensional CDT quantum gravity in the time 
direction -- if such a notion can be defined appropriately -- may be quantum Ricci-flat. 
It would be premature to draw any strong conclusions from our exploratory study, but the role of anisotropy in the curvature behaviour of CDT in
two dimensions clearly deserves further study.

\section{Summary and conclusion}
\label{sec:summary}

We set out to understand the curvature properties of the quantum spacetime generated dynamically in two-dimensional
CDT quantum gravity, working with standard boundary conditions, where both space and time are compactified
and the curved spacetimes summed over in the nonperturbative path integral therefore have the topology of a two-torus.
Using Monte Carlo simulations for discrete spacetime volumes of up to $N_2\! =\! 600k$, we measured average sphere distances
and the associated curvature profiles as functions of the link distance and the dual link distance. 
Our main objective was to distinguish between local and global features of the quantum Ricci curvature, and to isolate the former, to the
extent this is possible.
By ``global" we mean effects that are present purely as a consequence of the compactification. We saw that topological effects in the time direction are easily
taken care of by putting a sharp upper bound on $\delta$, which sets the linear size of the local neighbourhood that contributes to
a measurement of the average sphere distance, from which then the local quantum curvature is extracted.

A nontrivial part of the analysis dealt with establishing a similar bound to exclude the influence of the compactification in
the spatial direction. Here, large quantum fluctuations of the volume of the one-dimensional spatial slices prevent the existence of
a sharp bound. This motivated us to introduce a threshold criterion for the measured average sphere distances, which
involved the monitoring of what we called topological shortcuts. These are shortest geodesics between pairs of points on a double
circle, which wind around the torus in the ``wrong" way, in the sense that the geodesic would not exist if the local configuration was 
part of an infinitely extended planar triangulation. We classified a data point for an average sphere distance measurement --
contributing to the ensemble average -- as admissible if the fraction of point pairs 
connected by a topological shortcut did not exceed 1\%. From the statistics of these ``level-1 violations" we extracted
volume-dependent bounds $\delta_{\rm max}$, below which the measurements can be regarded as 
effectively free of topological effects.

Considering only data points that do not exceed these $\delta$-bounds
and therefore can be regarded as describing pure, quasi-local geometry, a single universal curvature profile 
$\langle \bar{d}_{\rm av}(\delta)/\delta \rangle$ is seen to emerge (yellow dots in Fig.\ \ref{fig:ric2dcdt-delmax}). Disregarding
lattice artefacts for $\delta\! \lesssim \! 5$, this curve increases monotonically throughout the investigated range 
$5\!\leq\! \delta\!\leq\! 40$, ostensibly indicating a small negative curvature.  
Assuming that the underlying quantum geometry is approximately homogeneous, a natural point of comparison is
a classical space of constant negative curvature. However, we dismissed this interpretation, since it would entail the presence of
a dynamically generated scale (a curvature radius), for which an obvious source is lacking. 
We were thus led to conclude that the curvature behaviour of the CDT quantum geometry falls into a new, nonclassical category, 
that of {\it quantum flatness}. The fact that no rescaling in $\delta$ seems necessary to obtain a common curvature profile indicates 
a scale-independent, fractal behaviour of the underlying quantum geometry, at least with respect to its (quasi-)local curvature properties. 
This does not contradict any
earlier findings on the geometric properties of two-dimensional CDT quantum gravity, which have largely concentrated on the dynamics of the spatial
volume, which is a {\it global} geometric quantity.\footnote{Our use of ``local"
and ``global" for geometric observables refers to the scale probed, in our case $\delta$. This is a meaningful distinction, even though in the
absence of other reference systems all observables include an averaging/integral over spacetime and therefore are always global (=nonlocal) 
in the standard classical sense.} 

Our result runs counter to the idea that the quantum-gravitating torus (on sufficiently coarse-grained scales and in the sense of expectation
values) should resemble a flat, classical torus in terms of its local curvature properties. 
Having finally a well-defined, nonperturbative notion of
renormalized curvature at our disposal \cite{qrc1,qrc2,qrc3,qrc4}, we have been able to test this idea, but it appears to be incorrect.  
Instead, on the basis of our measurements using both the link distance and the dual link distance and on theoretical grounds, we have concluded 
that the quantum geometry is characterised by a universal, non-constant curvature profile.
This behaviour, dubbed ``quantum flatness", appears to be a genuine nonperturbative feature of two-dimensional CDT quantum gravity on the torus,
rather than a numerical fluke. It would be interesting to better understand its origin in theoretical, analytical terms. 
Further input for this could come from a more detailed analysis of the anisotropic
features of the quantum Ricci curvature, extending our preliminary investigation in Sec.\ \ref{sec:anisotropy}.   
Another interesting question is whether quantum flatness is a phenomenon particular to two-dimensional Lorentzian quantum 
gravity or perhaps a more general feature of strongly fluctuating quantum geometry. A natural testing ground would be
fully-fledged CDT quantum gravity on a four-torus, to which our set-up and methodology naturally generalize, and which
has been at the focus of recent research \cite{toroidal}. Lastly, the present study provides further evidence for the quantum Ricci 
curvature as a powerful new ingredient to help us understand the nature of quantum geometry and quantum gravity.

\section*{Source code}
Our source code \cite{2dcdtgithub} for generating CDT configurations is open source, and freely available at \href{https://github.com/JorenB/2d-cdt}{github.com/JorenB/2d-cdt}.
\vspace{1cm}

\noindent {\bf Acknowledgments.} We thank G.\ Clemente for many useful discussions relating to Sec.\ \ref{sec:topology}. This work was partly supported by a
Projectruimte grant of the Foundation for Fundamental Research
on Matter (FOM, now defunct), financially supported by the Netherlands Organisation for Scientific Research (NWO).

\vspace{0.5cm}
\begin{appendices}

\section{}
\label{sec:appendix}
\begin{figure}[t]
	\centering
	\begin{subfigure}[t]{0.31\textwidth}
	\centering
		\includegraphics[width=0.9\textwidth]{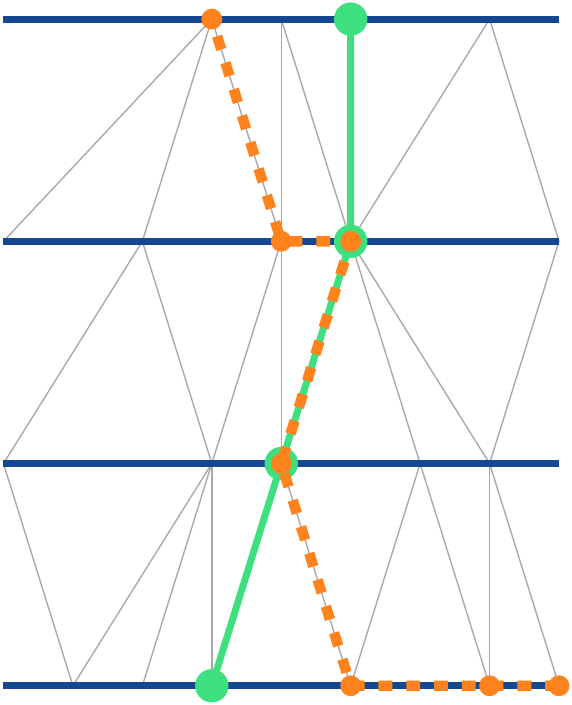}
		\caption{}
	\end{subfigure}
	\hfill
	\begin{subfigure}[t]{0.31\textwidth}
	\centering
		\includegraphics[width=0.9\textwidth]{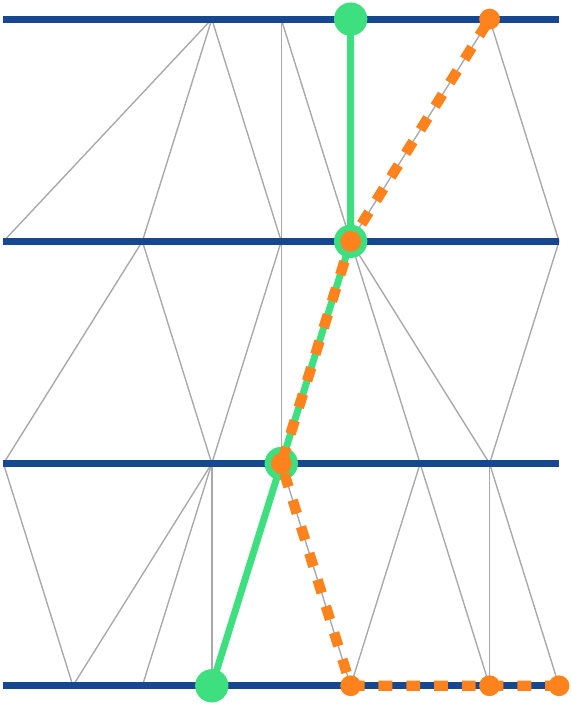}
		\caption{}
	\end{subfigure}
	\hfill
	\begin{subfigure}[t]{0.31\textwidth}
	\centering
		\includegraphics[width=0.9\textwidth]{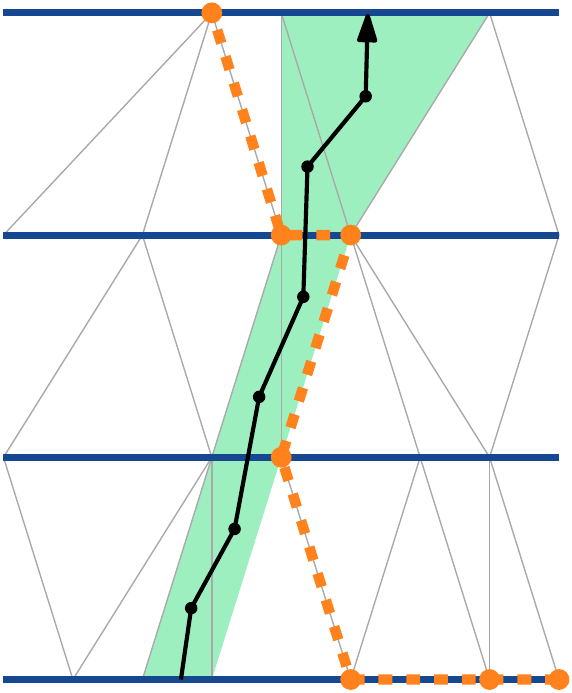}
		\caption{}
	\end{subfigure}
	\caption{
	Left, centre: if the reference curve is taken along the links of the triangulation (green), this leads to overlaps with other paths of the same type (orange, dotted) rather than point-like intersections.
	Right: if instead one of the paths is taken along dual links (black, dual to a contiguous ``chain" of triangles), intersections occur in isolated points only and the
	intersection number introduced in Sec.\ \ref{inter} is well defined.}
	\label{fig:crossing-primal-dual}
\end{figure}

In this appendix we explain how to set up the intersection number analysis introduced in Sec.\ \ref{inter} on CDT geometries, leading to the
results presented in Sec.\ \ref{measure}.\footnote{We will treat explicitly the case relevant for computing the curvature profile in terms of the link distance. There is
an analogous procedure if the dual link distance is used instead.}
As already mentioned earlier, this analysis is simplified when the reference loop $\gamma$ is constructed 
from dual links. Otherwise we would have to deal with situations where the two curves are ``tangent" in the sense of sharing one or more links, as
illustrated by Fig.\ \ref{fig:crossing-primal-dual}. If $\gamma$ consists of dual links, as in the black curve in Fig.\ \ref{fig:crossing-primal-dual}c, 
all intersection points occur between a link and a dual link, which always happens at a nonvanishing angle in an isolated point.  

For a given pair of circles on a CDT geometry $T$, we construct the curve $\gamma$ by randomly picking a triangle $\Delta$ outside the region $C$ 
enclosed by the two circles. From this starting point, we perform a breadth-first search (BFS) along dual links or, equivalently, along the triangles of $T$, 
again excluding all triangles located in $C$. The BFS is restricted to move forward in discrete CDT time $t$ or stay at constant time, but not go back in time.
The algorithm is continued until we meet the original triangle $\Delta$ again. We can then trace back our steps to find a shortest path connecting $\Delta$ to itself,
which will serve as the reference curve $\gamma$ (with winding number 1 in the positive time direction) for determining the intersection numbers with
all geodesics contributing to the average sphere distance computation. 
If the BFS exhausts the geometry while moving forward in time, without getting back to the triangle $\Delta$, we have encountered a level-2 violation, 
where the region $C$ wraps around the spatial direction in a full circle.

Once a curve $\gamma$ has been found, we enumerate all pairs of vertices $(q,q')\! \in\! S_p^\delta\! \times\! S_{p'}^\delta$,
and subsequently construct the shortest paths $\phi(q,q')$ connecting these pairs by again performing a BFS. Since we are interested
in their intersection number with $\gamma$, we not only keep track of the length of this shortest path, but also 
of all vertices encountered along the way. With this information, it is straightforward to compute the total intersection numbers $c(\phi,\gamma)$.

\begin{figure}[t]
	\centering
	\includegraphics[width=0.75\textwidth]{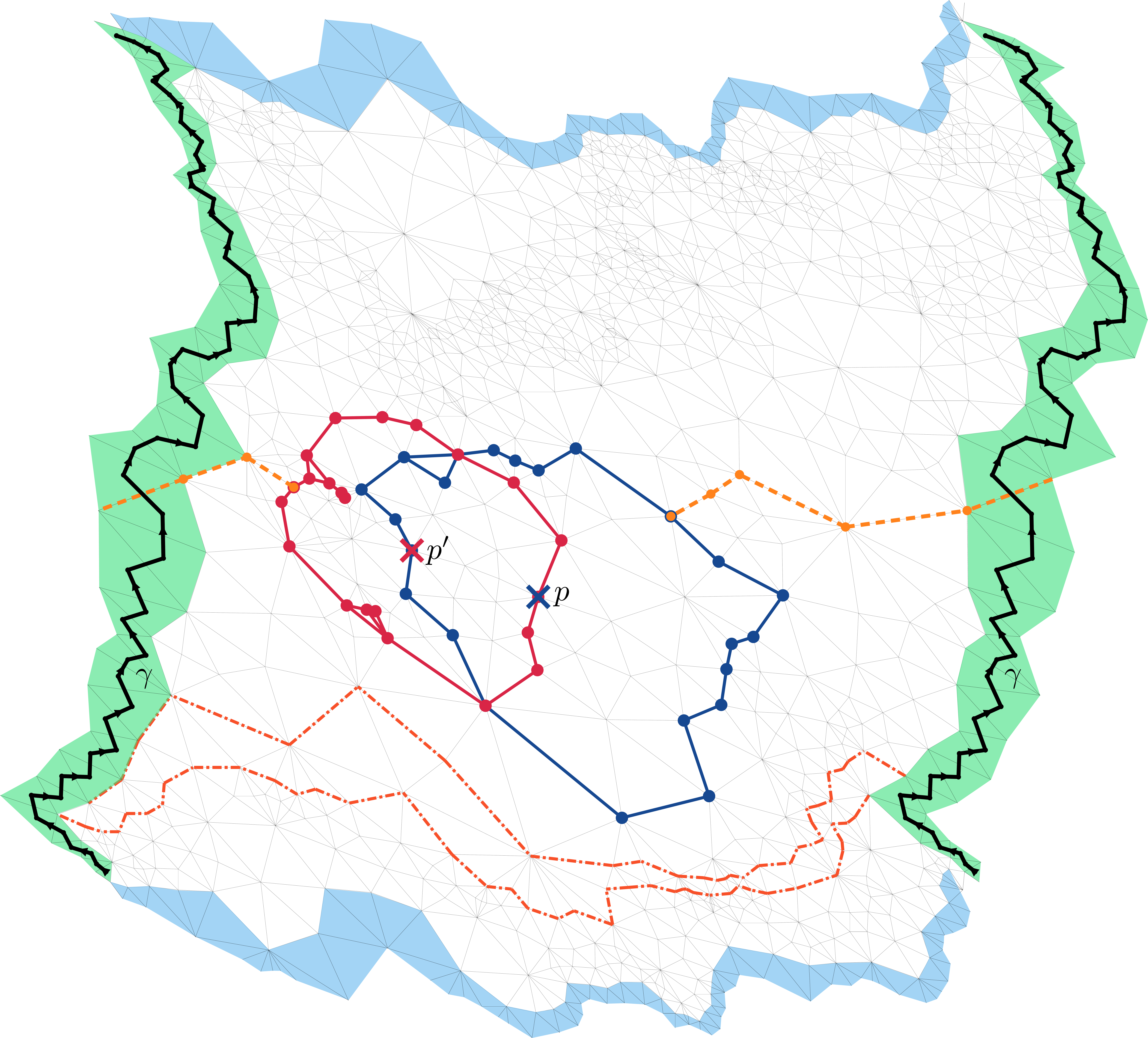}
	\caption{Example of a toroidal CDT triangulation, represented by a harmonic embedding. The centres of the circles $S_p^\delta$ (blue) 
	and $S_{p'}^\delta$ (red) of radius $\delta \!= \!3$ are marked by crosses. The presence of a shortest geodesic between two points 
	on the two circles (dashed, orange), which has nonzero intersection number with the reference curve $\gamma$ along dual links, signals a level-1 violation. 
	Opposite boundary strips should be identified, as described in the text. Two neighbouring spatial slices of constant time are indicated by 
	dash-dotted red lines.
	}
	\label{fig:triangulation-crossing-number}
\end{figure}

To illustrate some aspects of the nontrivial geometry of the CDT configurations on which this analysis takes place,  
Fig.\ \ref{fig:triangulation-crossing-number} depicts a specific example of such a triangulation. It contains a pair of circles 
$(S_p^3, S_{p'}^3)$ of radius 3, together with a pair of points connected by a shortest geodesic $\phi$ 
of nonvanishing intersection number. 
We have chosen to represent the triangulation by its harmonic embedding \cite{moduli} in the plane, 
which is based on two loops along its links that generate the fundamental group of the torus. They are a spacelike loop at constant time (bottom boundary in 
Fig.\ \ref{fig:triangulation-crossing-number}) and a timelike loop (left boundary in Fig.\ \ref{fig:triangulation-crossing-number}), which we have chosen to lie immediately to the left of the reference curve $\gamma$ used to compute the intersection numbers.
Opposite boundaries of this embedding should be identified. We have repeated the triangulated boundary strips (light blue) along opposite sides, which
makes it easier to see which boundary links should be identified pairwise to form the torus.

In addition to the two spheres and the topological shortcut $\phi$, which intersects the curve $\gamma$ exactly once, 
we have for reference also marked two neighbouring spatial slices (red, dash-dotted).
Unlike in the standard way of depicting two-dimensional CDT geometries (Fig.\ \ref{fig:cdt-sample}), in this representation the spatial slices
cannot be identified just by inspection. 
It is not coincidental that the circle configuration with a level-1 violation lies in a region that is relatively sparsely populated by triangles. 
Recall that (after the Wick rotation) all triangles of a CDT configuration are equilateral, a property that cannot be preserved faithfully by any
planar representation. The nature of the harmonic embedding we have chosen is such that regions with small spatial slices are stretched out, whereas regions with large spatial slices are squeezed. 
It implies that regions with lower triangle density are associated with smaller slice volumes, which are exactly the regions where we expect level-1 
violations to occur more frequently.

\end{appendices}

\end{document}

%% file: incs.tex
\usepackage{array,epsfig}
\usepackage{float}
\usepackage{amsmath}
\usepackage{amsfonts}
\usepackage{amssymb}
\usepackage{amsxtra}
\usepackage{amsthm}
\usepackage{dsfont}
\usepackage{mathrsfs}
\usepackage{pgf,tikz}
\usepackage{mathrsfs}
\usepackage{slashed}
\usepackage{mathtools}
\usepackage{color}
\usepackage{braket}
\usepackage{listings}
\usepackage{caption}
\usepackage{subcaption}
\usepackage{graphicx}
\usetikzlibrary{calc}
\usetikzlibrary{arrows}
\usetikzlibrary{plotmarks}

%% file: defs.tex
\theoremstyle{definition}

\newcommand{\bb}[1]{\mathbb{#1}}

\newcommand{\R}{\bb{R}}